\documentclass{jfmnomark}
\usepackage{graphicx}
\usepackage{natbib}
\usepackage{amsmath}
\usepackage{amscd}
\usepackage{pifont}
\usepackage{mathrsfs}
\usepackage{xfrac}
\usepackage{ifthen}
\usepackage{comment}

\DeclareMathAlphabet{\mathpzc}{OT1}{pzc}{m}{it}

\usepackage{tikz}

\tikzset{every mark/.style={scale=0.5, color=black}}

\tikzset{%
 dimengray/.style={<->,>=latex,thin,every rectangle node/.style={fill=gray!10,midway}},
  dimen/.style={<->,>=latex,thin,every rectangle node/.style={fill=white,midway}},
    every mark/.style={scale=0.5, color=black},
    every pin/.style={scale=0.8, rectangle, draw, thin, fill=gray!5, rounded corners=3pt, pin edge=thin, font=\footnotesize},
}

   \tikzstyle{dblpin}=[scale=1.0, fill=gray!5, rectangle split, rectangle split parts=2, rounded corners=3pt, font=\footnotesize]

   \tikzstyle{trippin}=[scale=1.0, fill=gray!5, rectangle split, rectangle split parts=3, rectangle split draw splits=false, rectangle split ignore empty parts, rounded corners=3pt, font=\footnotesize]

\tikzstyle{menode}=[scale=0.7, rectangle, draw, very thin, fill=gray!5, rounded corners=3pt, font=\footnotesize]
\tikzstyle{benode}=[scale=0.7, rectangle, draw, minimum width=2.2cm, very thin, fill=white, rounded corners=3pt, text=black, font=\footnotesize]

\newcommand*\mycircb[1]{%
  \tikz[baseline=-1.5pt]\node[draw,circle,inner sep=0.8pt, scale=0.6] 
  {\upshape \footnotesize \textbf{#1}};\!\!
}

\newcommand*\Swtch[3]{%
	\xrightarrow[\smash{{\text{\mycircb{#2} $>$ \mycircb{#3}}}}]{\text{#1}} 
}
\newcommand*\Swtchsym[3]{%
	\xrightarrow[\smash{\text{\mycircb{#2} $\gtrless$ \mycircb{#3}}}]{\text{#1}} 
}
\newcommand*\Swtchback[3]{%
	\xleftarrow[\smash{{\text{\mycircb{#2} $>$ \mycircb{#3}}}}]{\text{#1}} 
}
\newcommand*\Swtchsymback[3]{%
	\xleftarrow[\smash{\text{\mycircb{#2} $\gtrless$ \mycircb{#3}}}]{\text{#1}} 
}

\newcommand*\Mycirct[1]{%
  \tikz[baseline=-3pt] \draw[fill=white] (0,0) circle (4pt) node[scale=0.65] {\footnotesize \textbf{#1}};\!\!
}

\def\Xint#1{\mathchoice
   {\XXint\displaystyle\textstyle{#1}}%
   {\XXint\textstyle\scriptstyle{#1}}%
   {\XXint\scriptstyle\scriptscriptstyle{#1}}%
   {\XXint\scriptscriptstyle\scriptscriptstyle{#1}}%
   \!\int}
\def\XXint#1#2#3{{\setbox0=\hbox{$#1{#2#3}{\int}$}
     \vcenter{\hbox{$#2#3$}}\kern-.5\wd0}}

\def\dashint{\Xint-}

\usepackage{cases}

\ifCUPmtlplainloaded \else
  \checkfont{eurm10}
  \iffontfound
    \IfFileExists{upmath.sty}
      {\typeout{^^JFound AMS Euler Roman fonts on the system,
                   using the 'upmath' package.^^J}%
       \usepackage{upmath}}
      {\typeout{^^JFound AMS Euler Roman fonts on the system, but you
                   dont seem to have the}%
       \typeout{'upmath' package installed. JFM.cls can take advantage
                 of these fonts,^^Jif you use 'upmath' package.^^J}%
      }
  \else
  \fi
\fi

\ifCUPmtlplainloaded \else
  \checkfont{msam10}
  \iffontfound
    \IfFileExists{amssymb.sty}
      {\typeout{^^JFound AMS Symbol fonts on the system, using the
                'amssymb' package.^^J}%
       \usepackage{amssymb}%
         \let\leq=\leqslant
         \let\geq=\geqslant
      }{}
  \fi
\fi

\ifCUPmtlplainloaded \else
  \IfFileExists{amsbsy.sty}
    {\typeout{^^JFound the 'amsbsy' package on the system, using it.^^J}%
     \usepackage{amsbsy}}
    {}
\fi

\newsavebox{\astrutbox}
\sbox{\astrutbox}{\rule[-5pt]{0pt}{20pt}}

\title[New gravity-capillary waves (Part 2)]{New gravity-capillary waves at low speeds \\ Part 2: Nonlinear geometries}

\author[P.H. Trinh and S.J. Chapman]%
{P\ls H\ls I\ls L\ls I\ls P\ls P\ls E\ns H.\ns T\ls R\ls I\ls N\ls H$^{1,2}$\ns \and 
S.\ns J\ls O\ls N\ls A\ls T\ls H\ls A\ls N\ns C\ls H\ls A\ls P\ls M\ls A\ls
N$^2$ \break}

\affiliation{ $^1$Program in Applied and Computational Mathematics, Princeton University, \\ Washington Road,
Princeton, NJ, 08544, USA \\[\affilskip]
$^2$Oxford Centre for Industrial and Applied Mathematics, Mathematical Institute, \\ 24-29 St. Giles',
Oxford, Oxfordshire, OX1 3LB, UK}
\pubyear{1996}
\volume{538}
\pagerange{119--126}
\date{--- and in revised form ---}

\begin{document}

\maketitle

\begin{abstract}
When traditional linearised theory is used to study gravity-capillary waves produced by flow past an obstruction, the geometry of the object is assumed to be small in one or several of its dimensions. In order to preserve the nonlinear nature of the obstruction, asymptotic expansions in the low-Froude or low-Bond number limits can be derived, but here, the solutions are waveless to every order. This is because the waves are in fact, exponentially small, and thus \emph{beyond-all-orders} of regular asymptotics; their formation is a consequence of the divergence of the asymptotic series and the associated Stokes Phenomenon. 

In Part 1, we showed how exponential asymptotics could be used to study the problem when the size of the obstruction is first linearised. In this paper, we extend the analysis to the nonlinear problem, thus allowing the full geometry to be considered at leading order. When applied to the classic problem of flow over a step, our analysis reveals the existence of six classes of gravity-capillary waves, from which two share a connection with the usual linearised solutions first discovered by Lord Rayleigh. The new solutions arise due to the availability of multiple singularities in the geometry, coupled with the interplay of gravitational and cohesive effects.
\end{abstract}

\section{Introduction}

\noindent Consider water that flows past a fishing line, which we model as a pressure distribution applied near the surface. As \cite{rayleigh_1883} demonstrated, if the velocity of the stream is kept above some critical velocity, then capillary waves are produced upstream and gravity waves downstream. However, if the speed of the stream is too small, then subcritical or supercitical solitary waves are produced instead. This classification of the dynamics holds similarly for many flows past more general obstructions, where we can linearise the geometry in one or several of its dimensions. Thus, typical linearised theory does not really distinguish between flows past differently shaped objects, but merely requires that they are sufficiently small [see \emph{e.g.} \cite{forbes_1983} and \cite{king_1987}]. The question which we address in this paper is: \emph{what can be said about gravity-capillary flows over obstructions which are not linearised}?

There are two important parameters: the Froude number, $F$ (ratio of inertial to gravitational forces) and the Bond number, $B$ (ratio of gravitational to surface tension forces). In a previous work (\citealt{trinh_gclinear}---henceforth referred to as Part 1) we showed how the typical linearised theory can be re-interpreted in the low-Froude and low-Bond limits. If $q$ is the speed of the free-surface flow, and $F^2 = \mathcal{O}(\epsilon)$ while $B = \mathcal{O}(\epsilon^2)$, then the typical asymptotic expansion gives the \emph{base series}, which we denote as \Mycirct{B}, and write
\begin{equation} \label{eq:gcwave_genB} 
\text{\Mycirct{B}} \quad \Bigl[ q_0\, + \epsilon q_1 + \epsilon^2 q_2 + \mathcal{O}(\epsilon^3) \Bigr] e^0, 
\end{equation}

\noindent valid as $\epsilon \to 0$. The base series, \Mycirct{B}, is waveless to every order, and in fact, the gravity and capillary waves are exponentially small and thus \emph{beyond-all-orders}. These waves are \emph{switched-on} when \Mycirct{B} is analytically continued across critical curves (\emph{Stokes lines}) in the complex plane in a process known as the \emph{Stokes Phenomenon}. We use the notation \Mycirct{B} $>$ \Mycirct{G} or \Mycirct{B} $>$ \Mycirct{C} to indicate the switching-on of a gravity or capillary wave, respectively, and these waves are written as
\begin{equation} \label{eq:gcwave_genGC} 
\text{\Mycirct{G}} \quad \Bigl[ A_1 + \mathcal{O}(\epsilon)\Bigr] e^{-\chi_g/\epsilon} \qquad \text{and} \quad \qquad
\text{\Mycirct{C}} \quad \Bigl[ B_1 + \mathcal{O}(\epsilon)\Bigr] e^{-\chi_c/\epsilon}.
\end{equation}

\noindent In the linearised problem of flow over a step (Part 1), Stokes lines originate from the single, merged singularity which represents the step, and the point where the gravity and capillary waves are equal in magnitude corresponds to the critical bifurcation in the Froude-Bond plane.

The nonlinear analysis of this work differs from the linear analysis in two important ways. First, a nonlinear geometry usually contains multiple singularities and hence multiple Stokes lines; each singularity then has the potential to produce both gravity and capillary waves. Second, gravity and capillary waves can \emph{themselves} interact, with \Mycirct{G} $>$ \Mycirct{C} or \Mycirct{C} $>$ \Mycirct{G}. These more complicated secondary switchings may be accompanied by crossing Stokes lines, the higher-order Stokes Phenomenon, and other, more advanced, aspects of exponential asymptotics highlighted by \cite{howls_2004}, \cite{daalhuis_2004}, \cite{chapman_2005}, and \cite{chapman_2007}.

%This interaction of the three components, \mycirc{B}, \mycirc{G}, and \mycirc{C}, leads to a wider range of solution possibilities and 

Our main result shows that for a nonlinear obstruction, the typical bifurcation curve in the Froude-Bond plane \emph{thickens}, revealing new possibilities for solutions. The thickness of this curve is a manifestation of the finite nature of the obstruction, and shrinks to zero as the geometry is linearised. For example, in the case of a rectangular step, rather than the two standard linear solutions of \cite{rayleigh_1883}, there are now \emph{six} possible solutions, for which the previous two are special cases. 

\subsection{Physical and mathematical background}

A review of the physical and mathematical literature that motivates this work can be found in the introduction of Part 1 \citep{trinh_gclinear}. The situation of free-surface gravity-capillary waves produced by flow past an obstruction is a well studied problem, and so we shall only mention the more comprehensive reviews of the topic by \cite{dias_1999} and \cite{vb_book}, and of course, classic texts by \cite{lamb_book} and \cite{stoker_book}. 

As for the mathematical techniques we use in this paper known as \emph{exponential asymptotics}, these are directly based on the methods introduced by \cite{chapman_1998} and previously applied to study wave-structure interactions for pure capillary waves \citep{chapman_2002}, and pure gravity waves (\citealt{chapman_2006}, \citealt{trinh_1hull}, \citealt{lustri_2012}). A variety of other approaches to exponential asymptotics are available, and we refer the readers to the books by \cite{dingle_book}, \cite{costin_book}, and \cite{boyd_wnlsw} for an overview of the of the different techniques.

\section{Mathematical formulation}

\begin{figure} \centering 
\includegraphics[width=1.0\textwidth]{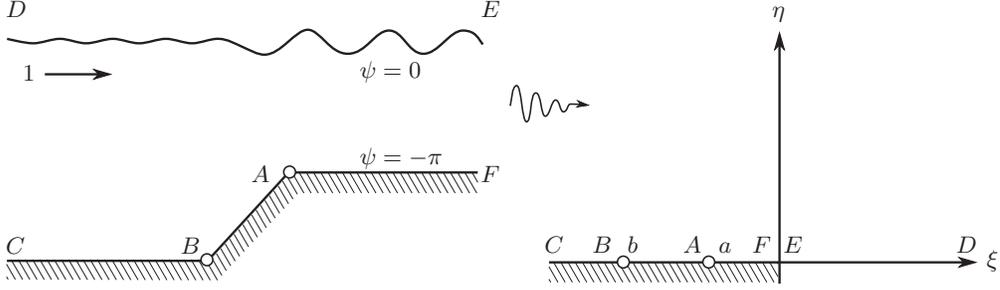}
\caption{We consider a typical flow over an obstruction. The flow in the
physical $xy$-plane (left) is first mapped to the potential $w = \phi + i\psi$
plane, then again mapped to the upper-half $\zeta = \xi + i\eta$ plane (right). For flow in a channel, the latter map is given by $\zeta = e^{-w}$. \label{fig:gcwave_form}}
\end{figure}

\noindent We briefly recapitulate the relevant equations, which parallel the ones presented in Part 1, but this time, we allow for the possibility of flows over more general geometries. Consider steady, two-dimensional potential flow of an incompressible fluid with upstream velocity $U$, and a prescribed length scale $L$. The flow is non-dimensionalised with characteristic scales of $U$ and $L/\pi$ for the velocity and lengths, respectively, and the physical $z = x + iy$ plane is mapped to the complex potential $w = \phi + i\psi$ plane, then mapped again to the upper-half $\zeta = \xi + i\eta$ plane using $\zeta = e^{-w}$. The free-surface is then given by
\begin{subequations}
\begin{gather}
 \log{q} = -\frac{1}{\pi}\dashint_{-\infty}^\infty
\frac{\theta(\xi')}{\xi'-\xi}
\ d\xi', \label{eq:gcwave_bdint_re} \\
 \beta \epsilon \left[ q^2 \frac{dq}{d\phi} \right]
- \beta \tau \epsilon^2 \left[ q^2 \frac{d^2\theta}{d\phi^2} +
 q\frac{dq}{d\phi}\frac{d\theta}{d\phi} \right] =
- \sin\theta \label{eq:gcwave_dyn_re},
\end{gather}
\end{subequations}
 
 \noindent where $F^2 = \beta \epsilon$ is the square of the Froude number, 
$B = \beta\tau \epsilon^2$ is the Bond number, and for hodograph variables
$\log (dw/dz) = \log q - i\theta$, where $q$ is the fluid speed and $\theta$ is the
angle the streamlines make with the $x$-axis. This set-up is presented in
Figure \ref{fig:gcwave_form}. 

Our analysis must then be extended to the complexification of the free-surface (originally, $q$ and $\theta$ for $\xi \in \mathbb{R}^+$). Complexifying the free-surface into the upper-half $\xi$-plane and relabeling $\xi \mapsto \zeta$ and $\phi \mapsto w$ gives
\begin{subequations}
\begin{gather} 
 \log{q} - i\theta =   -\frac{1}{\pi}\int_{-\infty}^0
\frac{\theta(\xi')}{\xi'-\zeta} \ d\xi' +  \mathscr{H}\theta(\zeta)
\label{eq:gcwave_bdint}, \\
 \beta \epsilon \left[ q^2 \frac{dq}{dw} \right]
- \beta\tau \epsilon^2 \left[ q^2 \frac{d^2\theta}{dw^2} +
 q\frac{dq}{dw}\frac{d\theta}{dw} \right] =
- \sin\theta. \label{eq:gcwave_dyn}
\end{gather} 
\end{subequations}

\noindent where $\mathscr{H}$ denotes the Hilbert transform operator on the
semi-infinite interval $\xi \geq 0$ which corresponds to the free-boundary, \emph{i.e.}
 \begin{equation}
 \mathscr{H}\theta(\zeta) =  -\frac{1}{\pi}\int_{0}^\infty
\frac{\theta(\xi')}{\xi'-\zeta} \ d\xi'.
 \end{equation}

\subsection{The inclined step}

\noindent The methodology developed throughout this paper is applicable to
most general non-surface-piercing geometries, which are chosen by imposing the
value of $\theta$ along the negative $\zeta = \xi$ axis. With minimal modification, it
can also be applied to surface-piercing obstructions (such as for a ship), the
only significant difficulty being the assumptions-to-make near the point of
contact [see \cite{trinh_1hull}]. 

However, for concreteness and illustration, we will often take as an example the
step in a channel, with 
\begin{equation} \label{wzeta}
\zeta = e^{-w}
\end{equation}

\noindent and
\begin{equation} \label{eq:gcwave_thetageom}
\theta = \begin{cases} 
0 & \text{for $\zeta < -b$} \\
\sigma\pi & \text{for $\zeta \in (-b, -a)$} \\
0 & \text{for $\zeta \in (-a, 0)$}
\end{cases}
\end{equation}

\noindent where $0 < a < b$. We will always imagine a stepping \emph{up} (from
left-to-right), with $b$ associated with the stagnation point, $a$ associated
with the corner, and therefore $0 < \sigma < 1$. In the next section,
we will see that at zero Froude and Bond numbers, the leading-order solution is
provided by substituting the above values for $\theta$ along $\zeta < 0$, as well as  $\theta = 0$ for $\zeta > 0$, into the boundary integral equation \eqref{eq:gcwave_bdint}, giving
\begin{equation} \label{eq:gcwave_rigidwall}
q \sim \left(\frac{\zeta + b}{\zeta + a}\right)^{\sigma} \text{\qquad and
\qquad} \theta = 0 + \mathcal{O}(\epsilon),
\end{equation}

\noindent which is often termed the \emph{rigid-wall flow}, as it is equivalent to replacing the free surface by the rigid wall, $\theta = 0$.

\section{Asymptotic approximation} \label{sec:gcwave_asym}

\noindent Substituting the usual perturbation expansions (the base series), 
\begin{equation} \label{eq:gcwave_asymseries}
 \theta \sim \sum_{n=0}^\infty \epsilon^n \theta_n 
 \text{\ \quad and \ \quad}
 q \sim \sum_{n=0}^\infty \epsilon^n q_n
\end{equation}

\noindent into \eqref{eq:gcwave_bdint} and \eqref{eq:gcwave_dyn}
yields at
$\mathcal{O}(1)$, 
\begin{subequations}
\begin{align}
\theta_0	& = 0, \label{eq:gcwave_theta0} \\
\log{q_0} 	& = -\frac{1}{\pi}\int_{-\infty}^0 \frac{\theta(\xi')}{\xi'-\zeta} \ d\xi' \label{eq:gcwave_logq0}, \\
\intertext{\noindent and at $\mathcal{O}(\epsilon)$,} 
q_0^2 \frac{dq_0}{dw} &= -\frac{1}{\beta} \theta_1,  \label{eq:gcwave_theta1} \\
 \frac{q_1}{q_0}-i\theta_1 	&= 	\mathscr{H}\theta_1(\zeta). \label{eq:gcwave_q1}
\end{align}
\end{subequations}

\noindent Notice that imposing the bottom topography in \eqref{eq:gcwave_logq0} determines $\theta$ along $\xi \in \mathbb{R}^-$. The leading order solutions \eqref{eq:gcwave_theta0}--\eqref{eq:gcwave_logq0}, simply correspond to the rigid wall flow whereby the free-surface is replaced by $\theta = 0$. 

The full expressions for the higher $\mathcal{O}(\epsilon^n)$ terms are prohibitively complicated, but since we are mainly concerned with the limit $n \to \infty$, we may proceed in the following manner: the complexification of the leading-order free surface, $q_0$ will typically contain singularities, identifiable with singularities in the flow-domain, such as those corresponding to corners or stagnation points. However, because all the higher-order problems are linear, no new singularities can be introduced and thus, for all $n$, the singular points of $q_n$ must be those \emph{same} points as for $q_0$. 

Now if we examine the dynamic condition \eqref{eq:gcwave_dyn}, we can see that each successive term of the asymptotic approximation requires the derivative of the previous term. Then if $q_n$ contains a singularity of the form $1/(w-w^*)^n$, $q_{n+1}$ will contain a singularity of the form $n/(w-w^*)^{n+1}$. Thus as $n\to \infty$, we can expect the late terms to behave like factorial over power, or
\begin{equation} \label{eq:gcwave_ansatz}
 \theta_n \sim \frac{\Theta(w)\Gamma(n+\gamma)}{\chi(w)^{n+\gamma}}
 \text{\ \quad and \ \quad}
 q_n \sim \frac{Q(w)\Gamma(n+\gamma)}{\chi(w)^{n+\gamma}}.
\end{equation}

In fact, the inductive argument used to explain the divergence of the asymptotic expansion according to \eqref{eq:gcwave_ansatz} also serves to explain why the free surface waves are expected to be exponentially small in the limit $\epsilon \to 0$: the leading order solution \eqref{eq:gcwave_rigidwall} is waveless, and since the values of $q_n$ and $\theta_n$ only depend on the derivatives of the previous orders, then none of the terms in the naive expansion \eqref{eq:gcwave_asymseries} will contain waves. 

Proceeding then with the ansatz \eqref{eq:gcwave_ansatz}, we can now pinpoint the necessary terms required at $\mathcal{O}(\epsilon^n$). In the limit that $n\to \infty$, terms like $q_m
q_{n}$ for $m$ finite dominate terms with smaller indices in $n$, such as $q_m q_{n-1}$. Moreover, differentiating a term increases the order (in $n$) by 1, so a term
like $\epsilon dq_{n-1}/dw$ is of the same order as $q_n$. The relevant terms at $\mathcal{O}(\epsilon^n)$ of the boundary
integral equation \eqref{eq:gcwave_bdint} are thus
\[
 \frac{q_n}{q_0} - \frac{q_{n-1}q_1}{q_0^2} + \ldots - i\theta_n =
\mathscr{H}\theta_n(\zeta),
\]

\noindent for $n \geq 2$. It is known that as $n\to\infty$, the Hilbert transform on the right-hand side of this equation is exponentially subdominant to the terms on the left [see Part 1 or \emph{e.g.} \cite{chapman_2006} for further details]. Thus,
\begin{equation}
 \theta_n \sim -i\frac{q_n}{q_0} + \frac{iq_1 q_{n-1}}{q_0^2} + \ldots \quad
\textrm{as } n \to \infty, 
\end{equation}

\noindent or, 
\begin{equation} \label{eq:gcwave_qt2}
 q_n \sim iq_0 \theta_n + i\theta_{n-1}q_1 + \ldots \quad \textrm{as } n \to
\infty.
\end{equation}

\noindent As for Bernoulli's Equation \eqref{eq:gcwave_dyn}, we will use \eqref{eq:gcwave_qt2} to replace $q_n$ with $\theta_n$,
after which the relevant terms at $\mathcal{O}(\epsilon^n)$ are
\begin{multline} \label{eq:gcwave_dynleadn}
\overbrace{\Bigl[iq_0^3 \Bigr] \theta_{n-1}' - \Bigl[ \tau q_0^2 \Bigr]
\theta_{n-2}'' + \Bigl[ \frac{1}{\beta} \Bigr] \theta_n}^\text{first and
second order as $n\to\infty$} \\
+ \underbrace{\Bigl[ 3i q_0^2 q_0' \Bigr] \theta_{n-1} + \Bigl[ 3iq_0^2 q_1 -
\tau q_0 q_0' \Bigr] \theta_{n-2}' - \Bigl[ 2\tau q_0 q_1
\Bigr] \theta_{n-3}''}_\text{second order as $n\to\infty$} = 0,
\end{multline}

\noindent with primes for differentiation in $w$. Using the ansatz \eqref{eq:gcwave_ansatz} in the above equation, we have at leading order as $n\to\infty$,
\begin{equation} \label{eq:gcwave_chiquadratic}
 -i q_0^3 \chi' - \tau q_0^2 (\chi')^2 + \frac{1}{\beta} = 0,
\end{equation}

\noindent which is simply solved to give
\begin{equation} \label{eq:gcwave_chiprime}
 \frac{d\chi}{dw} = -i \left[ \frac{q_0^2 \pm \sqrt{\Delta}}{2\tau q_0}
\right],
\end{equation}

\noindent where
\begin{equation} \label{eq:gcwave_DeltaA}
\Delta = q_0^4 - A \text{\qquad and \qquad} A = \frac{4\tau}{\beta}, 
\end{equation} 

\noindent and $A$ will turn out to be a key parameter. Remember that $\chi$ is the portion of the ansatz \eqref{eq:gcwave_ansatz} that expresses the singularities of the higher order terms. Since $\chi(w^*) = 0$ for one of these singularities, $w^*$, we may express
\begin{equation} \label{eq:gcwave_chi}
\chi_\pm(w) = -i \int_{w^*}^w \left[ \frac{q_0^2 \pm
\sqrt{\Delta}}{2\tau q_0} \right] \ d\varphi.
\end{equation}

\noindent Note that in writing \eqref{eq:gcwave_chi}, we shall restrict the path of integration to be along \emph{the same Riemann sheet} as $w^*$. Integration contours that traverse different sheets (for example, by crossing the branch cut(s) from $\Delta = 0$) will be addressed in later sections, and in particular, Appendix \ref{sec:crossing}. By taking the limit of \eqref{eq:gcwave_chi} as $\tau \to 0$ and comparing with Chapman and Vanden-Broeck \citeyearpar{chapman_2002, chapman_2006}, we see that the positive sign corresponds to capillary waves and the negative sign to gravity waves. 

%We introduce an alternative notation which serves to clarify the labeling issues associated with the various Riemann sheets of \eqref{eq:gcwave_chiprime} and \eqref{eq:gcwave_chi} [see the Appendix of \cite{chapman_rupture} for a similar notational usage]. Rather than using the $\pm$ signs, which will need to be re-labelled if the integration contour in \eqref{eq:gcwave_chi} crosses a branch cut, we let $w = w^*_{(k)}$ correspond to a point on the $k^\text{th}$ Riemann sheet associated with $\sqrt{\Delta}$. Thus, $w^*_{(k)}$ is mapped to $e^{\pi i k}\sqrt{\Delta}$, with $k = 0, 1$. We then write $\chi$ as 
%\begin{equation} \label{eq:gcwave_chi2}
% \chi_{k} = -i \int_{w^*_{(k)}}^w \left[ \frac{q_0^2 + e^{\pi i k(s)} \sqrt{\Delta}}{2\tau q_0} \right] \ d\varphi,
%\end{equation}
%
%\noindent where $k(\varphi) = 0$ or $1$, depending on which sheet the integrand is currently on. The advantage of using \eqref{eq:gcwave_chi2} is that we may continue to associate, for example, a capillary wave with the singularity $w^*_{(0)}$ and initial branch $\sqrt{\Delta}$, even though the value of $w$ in the upper limit may have crossed onto the second Riemann sheet. In cases where only local values of $\chi$ is required, we shall use \eqref{eq:gcwave_chi}, but in cases where the global values of $\chi$ are desired, we prefer to work with \eqref{eq:gcwave_chi2}.

Now returning to the dynamic condition \eqref{eq:gcwave_dynleadn} and with the ansatzes \eqref{eq:gcwave_ansatz}, we find at next order in $n$: 
\begin{multline} 
\Bigl[iq_0^3 \Bigr] \Theta' - 
\Bigl[ \tau q_0^2 \Bigr] \Bigl\{-2\chi' \Theta' - \chi''\Theta\Bigr\} +
\Bigl[ 3i q_0^2 q_0' \Bigr] \Theta \\ +
\Bigl[ 3iq_0^2 q_1 - \tau q_0 q_0' \Bigr]\Bigl\{-\chi' \Theta\Bigr\}  -
\Bigl[ 2\tau q_0 q_1 \Bigr]\Bigl\{(\chi')^2 \Theta\Bigr\} = 0.
\end{multline}

\noindent We may write this as 
\begin{equation} \label{eq:gcwave_Thetatmp}
 \frac{\Theta'}{\Theta} = -\frac{1}{4}\frac{\Delta'}{\Delta} \pm 
 \frac{\Delta'}{2\sqrt{\Delta(\Delta+A)}} + F_\pm(w),
\end{equation}

\noindent with 
\begin{equation}
 F_\pm(w) \equiv \frac{iq_1}{2\tau} \left[ 1 \pm \frac{\Delta + 2A}{\sqrt{\Delta(\Delta + A)}} \right],
\end{equation}

\noindent and thus, after one integration,
\begin{equation} \label{eq:gcwave_Theta}
 \Theta(w) = \frac{\bigl(\sqrt{\Delta+A} + 
 \sqrt{\Delta}\bigr)^{\pm  1}\Lambda_\pm}{\Delta^{1/4}} \times \exp \left[
 \int_{w^\bigstar}^w F_\pm(\varphi) \ d\varphi \right],
\end{equation}

\noindent where $\Lambda_\pm$ is a constant and we may begin the integration at any arbitrary point $w = w^\bigstar$ where the integral is defined (often $w^*$ is a natural choice for $w^\bigstar$, but the integral may not exist at $w^*$). Note that \eqref{eq:gcwave_qt2},
allows us to relate $Q$ and $\Theta$, with
\begin{equation} \label{eq:gcwave_Q}
 Q \sim iq_0 \Theta.
\end{equation}

\noindent As a check, we may take the limit of $\tau \to 0$ and use the negative sign of \eqref{eq:gcwave_Theta}. This recovers the pre-factor for the gravity-wave problem,
\begin{equation} \label{Theta}
\Theta_{-}(w) = -\frac{\Lambda_- i}{2q_0^3} \exp \left[ -3i\int_{w^\bigstar}^w \frac{q_1}{q_0^4} \ d\varphi \right],
\end{equation}

\noindent to leading order, which differs from (3.16) in \cite{chapman_2006} by a factor of a half (our $\Lambda_-$ is their $2\Lambda$). 

\subsection*{A note on the $\pm$ notation}

As we have seen, each singularity is associated with two branches of $\chi = \chi_\pm$ in \eqref{eq:gcwave_chi}, for which the positive sign has been shown to correspond to capillary waves and the negative sign to gravity waves. There are also associated $\Theta = \Theta_\pm$, $Q = Q_\pm$, and $\Lambda = \Lambda_\pm$ quantities. In what follows,  we will sometimes use non-subscripted variables if it is unimportant which branch of $\chi$ we are referring to. 

\section{Optimal truncation and Stokes line smoothing} \label{sec:gcwave_smooth}

\noindent The underlying divergence of the asymptotic expansions will cause the \emph{Stokes Phenomenon} to occur: as the complexified asymptotic solution crosses a critical line (the \emph{Stokes line}), a small exponential switches on. Because the switching-on of the exponential is almost always via an error function \citep{berry_1989}, the optimal truncation and Stokes smoothing procedure will be similar to the one we performed in Part 1, with the exception of the nonlinear terms. 

To begin, we \emph{truncate} the asymptotic series at $n = \mathcal{N}$ so that
\begin{equation} \label{eq:gcwave_trun}
\theta = \sum_{n=0}^{\mathcal{N}-1} \epsilon^n \theta_n + R_\mathcal{N} 
\qquad \text{and} \qquad
q = \sum_{n=0}^{\mathcal{N}-1} \epsilon^n q_n + S_\mathcal{N},
\end{equation}

\noindent where the remainders are related by \eqref{eq:gcwave_qt2}
and thus
\begin{equation} \label{eq:gcwave_SnRn}
\frac{S_\mathcal{N}}{q_0}  - \frac{\epsilon q_1 S_\mathcal{N}}{q_0^2} - \frac{\epsilon^\mathcal{N} q_1
q_{\mathcal{N}-1}}{q_0^2} + \ldots = iR_\mathcal{N}.
\end{equation}
 
\noindent We will substitute the truncated sums \eqref{eq:gcwave_trun} into
Bernoulli's Equation \eqref{eq:gcwave_dyn} and in doing so, we will see two
separate types of terms, the ones involving only $q_n$ and $\theta_n$, and the ones involving the remainders, $R_\mathcal{N}$ and $S_\mathcal{N}$. 

Let us first study the terms involving the remainders. The remainder $S_\mathcal{N}$ can be written in terms of $R_\mathcal{N}$ by \eqref{eq:gcwave_SnRn} and after making this substitution, we are particularly interested in the leading-order terms, which are indicated by factors of $R_\mathcal{N}, \epsilon R_\mathcal{N}'$, and $\epsilon^2 R_\mathcal{N}''$, and second-order terms, which are indicated by factors of $\epsilon R_\mathcal{N}$, $\epsilon^2 R_\mathcal{N}'$, and $\epsilon^3 R_\mathcal{N}''$. The relevant terms from Bernoulli's Equation are
then given by the linear operator defined by
\begin{multline} \label{eq:gcwave_RNeq}
\mathcal{L}(R_\mathcal{N}; \epsilon) = 
\Bigl[iq_0^3 \Bigr] \epsilon R_\mathcal{N}' - \Bigl[ \tau q_0^2 \Bigr]
\epsilon^2 R_\mathcal{N}'' + \Bigl[ \frac{1}{\beta} \Bigr] R_\mathcal{N} \\
+ \Bigl[ 3i q_0^2 q_0' \Bigr] \epsilon R_\mathcal{N} + \Bigl[ 3iq_0^2 q_1 -
\tau q_0 q_0' \Bigr] \epsilon^2 R_\mathcal{N}' - \Bigl[ 2\tau q_0 q_1
\Bigr] \epsilon^3 R_\mathcal{N}'',
\end{multline}
 
\noindent which is exactly the same form as the left of \eqref{eq:gcwave_dynleadn}. We then introduce the Stokes smoothing parameter, $\mathcal{S} = \mathcal{S}(w)$, and set $R_\mathcal{N} = \mathcal{S} [\Theta e^{-\chi/\epsilon}]$. We know that the ansatz $\Theta e^{-\chi/\epsilon}$ solves $\mathcal{L} = 0$, so only terms involving derivatives of $\mathcal{S}$ will be left. After some computation, we find to leading order, 
\begin{align*}
 \mathcal{L}(R_\mathcal{N}; \epsilon) &\sim \epsilon \Theta e^{-\chi/\epsilon}
\frac{d\mathcal{S}}{dw} \Bigl[ iq_0^3 + 2\tau q_0^2 \chi' \Bigr]. \\
\intertext{Writing $d\mathcal{S}/dw = \chi' d\mathcal{S}/d\chi$, and using \eqref{eq:gcwave_chiquadratic} gives}
\mathcal{L}(R_\mathcal{N}; \epsilon) &\sim \epsilon \Theta e^{-\chi/\epsilon}
\frac{d\mathcal{S}}{d\chi} \Bigl[ \frac{1}{\beta} + \tau q_0^2 (\chi')^2 \Bigr].
\end{align*}

\noindent Now let us turn to the terms involving $q_n$ and $\theta_n$: when the
truncated sum in \eqref{eq:gcwave_trun} is substituted into \eqref{eq:gcwave_dyn}, terms of $\mathcal{O}(\epsilon^{\mathcal{N}-1})$ will automatically
be satisfied, and this process leaves us only with the remnant
$\mathcal{O}(\epsilon^{\mathcal{N}})$ contributions from the inertial terms, as well as the $\mathcal{O}(\epsilon^\mathcal{N})$ and $\mathcal{O}(\epsilon^{\mathcal{N}+1})$ contributions from the surface-tension terms:
 \begin{multline} \label{eq:gcwave_truntmp}
\epsilon^\mathcal{N} \left[ q_0^2 \frac{dq_{\mathcal{N}-1}}{dw} + \ldots \right] 
- \epsilon^{\mathcal{N}} \tau \left[ q_0^2 \frac{d^2\theta_{\mathcal{N}-2}}{dw^2} + \ldots \right]
- \epsilon^{\mathcal{N}+1} \tau \left[ q_0^2 \frac{d^2\theta_{\mathcal{N}-1}}{dw^2} + \ldots \right]
\\ 
= \epsilon^{\mathcal{N}} \biggl[ -\frac{\theta_\mathcal{N}}{\beta} + \ldots \biggr]
-\epsilon^{\mathcal{N}+1} \tau \left[ q_0^2 \frac{d^2\theta_{\mathcal{N}-1}}{dw^2} + \ldots \right],
\end{multline}

\noindent and thus in total we have
\begin{equation} \label{eq:gcwave_Lop}
\mathcal{L}(R_\mathcal{N}; \epsilon) \sim \epsilon^\mathcal{N}
\left[\frac{\theta_\mathcal{N}}{\beta}+ \epsilon \tau q_0^2 \frac{d^2\theta_{\mathcal{N}-1}}{dw^2}
\right].
\end{equation}

\noindent Since we are interested in the limit $\epsilon \to 0$ when $\mathcal{N} \to
\infty$, we can substitute the late-orders ansatz of \eqref{eq:gcwave_ansatz}
into the right-hand side of \eqref{eq:gcwave_Lop}, giving
\begin{equation} \label{eq:gcwave_Lop_rhs}
\mathcal{L}(R_\mathcal{N}; \epsilon) 
\sim \frac{\epsilon^{\mathcal{N}} 
\Theta \Gamma(\mathcal{N}+\gamma)}{\chi^{\mathcal{N}+\gamma}} 
\left[ \frac{1}{\beta} + \tau q_0^2 (\chi')^2 
\frac{\epsilon(\mathcal{N}+\gamma+1)}{\chi} \right]. 
\end{equation}

\noindent We introduce a coordinate system along the Stokes line using $\chi = re^{i\vartheta}$. The optimal truncation point (where adjacent terms of the expansion are equal in size) is at $\lceil r/\epsilon \rceil$, so we write $\mathcal{N} = r/\epsilon + \rho$ where $\rho \in [0, 1)$.  Changing to differentiation in $\vartheta$, using 
\[
 \frac{d}{d\chi} = -i\frac{e^{-i\vartheta}}{r} \frac{d}{d\vartheta},
\]

\noindent and applying Stirling's formula to \eqref{eq:gcwave_Lop_rhs} gives
\[
\frac{d\mathcal{S}}{d\vartheta} \left[ \frac{1}{\beta} + \tau q_0^2 (\chi')^2 \right]
\sim 
\frac{\sqrt{2\pi r}}{\epsilon^{\gamma+1/2}}
\left[ \frac{1}{\beta} + \frac{\tau q_0^2 (\chi')^2}{e^{i\theta}} \right]
\left(e^{-i\vartheta}\right)^{r/\epsilon+\rho+\gamma}
e^{re^{i\vartheta}/\epsilon} e^{-r/\epsilon}.
\]

\noindent The exponential factor on the right is exponentially small, except
near the Stokes line $\vartheta = 0$, where the critical scaling occurs with
$\vartheta = \sqrt{\epsilon}\overline{\vartheta}$. In total, then,
\[
\frac{d\mathcal{S}}{d\overline{\vartheta}} \sim \frac{i \sqrt{2\pi r}}{\epsilon^\gamma}
\exp \left( -\frac{r\overline{\vartheta}^2}{2} \right),
\]

\noindent and we have recovered the typical error function expression for the Stokes multiplier, $\mathcal{S}$. We now integrate this expression across the Stokes line from upstream, or $\overline{\vartheta} = \infty$, to downstream, or $\overline{\vartheta} = -\infty$. The apparent jump in the remainders of $\theta$ and $q$ are then
\begin{subequations}
\begin{align} 
\Bigl[ R_\mathcal{N} \Bigr]_\text{upstream}^\text{downstream} &\sim -\frac{2\pi i}{\epsilon^\gamma} \Theta e^{-\chi/\epsilon} \equiv \theta_\text{exp}, \label{eq:gcwave_stokes_jumpt} \\
\Bigl[ S_\mathcal{N} \Bigr]_\text{upstream}^\text{downstream} &\sim -\frac{2\pi i}{\epsilon^\gamma} Qe^{-\chi/\epsilon} \equiv q_\text{exp} \label{eq:gcwave_stokes_jumpq}
\end{align}
\end{subequations}

%\noindent with a similar exponentially small correction to $q$ given by

\noindent We can re-introduce the notation for the switching-on mechanism used in Part 1. We write, for example,
\begin{align*}
\sum_{n} \epsilon^n q_n &\Swtch{Stag.}{B}{C} 
\sum_{n} \epsilon^n q_n + q_\text{exp} \\
\sum_{n} \epsilon^n q_n &\Swtch{Corn.}{B}{G} 
\sum_{n} \epsilon^n q_n + q_\text{exp}, 
\end{align*}

\noindent and the arrow notation should be read as ``the base series turns on a capillary/gravity wave as the Stokes line from the stagnation/corner point is crossed''. 

To finalise the analysis, we need to also complexify the free boundary into the lower-half plane. This analogous process yields the functional complex conjugates of \eqref{eq:gcwave_stokes_jumpt} and \eqref{eq:gcwave_stokes_jumpq}, and thus the total leading order contribution along the
free-surface is given by twice the real parts of \eqref{eq:gcwave_stokes_jumpt} and \eqref{eq:gcwave_stokes_jumpq}, 
\begin{subequations} \label{qexpthetaexp}
\begin{align}
q_\text{exp, total} &\sim -\frac{4\pi}{\epsilon^\gamma} \Im \left(Qe^{-\chi/\epsilon}
\right), \label{eq:gcwave_qexp} \\
\theta_\text{exp, total} &\sim -\frac{4\pi}{\epsilon^\gamma} \Im
\left(\Theta e^{-\chi/\epsilon} \right). \label{eq:gcwave_thetaexp} 
\end{align} 
\end{subequations}

In order to fully determine the waves, we must also compute the values of the pre-factor, $\Lambda$, that appears in \eqref{eq:gcwave_Q} and \eqref{Theta} for $Q$ and $\Theta$, and also the value of $\gamma$ in \eqref{qexpthetaexp}. This can be done by matching the outer solution \eqref{eq:gcwave_asymseries} with an inner solution valid near each of the singularities. Although this was easily done for the linear problem in Part 1, the nonlinear problem is much more tedious, and readers can refer to \cite{trinh_thesis} for the detailed procedure.  

%Remember also that the exponentially small quantities in \eqref{eq:gcwave_stokes_jumpt}--\eqref{eq:gcwave_thetaexp} are only the leading order terms of a series in powers of $\epsilon$ [\emph{i.e.} the quantities $A_1$ and $B_1$ in \eqref{eq:gcwave_genGC}].

\section{Inner limits of $\chi$ and Stokes lines} \label{sec:gcwave_chistokes}

\noindent In the low-Froude and low-Bond limits, we can think of the exponentially small free-surface waves as having been generated by singularities in the flow domain. For example, in the case of the step \eqref{eq:gcwave_rigidwall}, both the corner and stagnation points are responsible for producing waves through their associated Stokes lines. Furthermore, we know from \cite{dingle_book} that Stokes lines with \Mycirct{B} $>$ \Mycirct{C} or \Mycirct{B} $>$ \Mycirct{G}, are given at points where the base series (with phase zero) reaches peak exponential dominance over the capillary or gravity waves (with phase $-\chi/\epsilon$). This yields the two conditions
\begin{equation} \label{eq:gcwave_chistokes} 
\Im(\chi_\pm) = 0 \qquad \text{and} \qquad \Re(\chi_\pm) \geq 0,
\end{equation}

\noindent and for a given geometry, the trajectory of the Stokes lines can be derived by numerically evaluating the integral \eqref{eq:gcwave_chi} and applying conditions \eqref{eq:gcwave_chistokes}. Whether a Stokes line encounters the free surface is a problem that thus depends on the global behaviour of $q_0$; however, we can still study the local properties of the Stokes lines near their associated singularities.

\subsection{Local Stokes line analysis for general channel flows} 

Let us assume that there is a singularity in the outer solutions \eqref{eq:gcwave_asymseries} at the points $z = z^*$, $w = w^*$, and $\zeta = \zeta^*$, which correspond to the different planes introduced in Figure \ref{fig:gcwave_form}. We introduce the shifted coordinate $W = w - w^*$, and then from potential theory, $W \sim \textrm{const} \times (z-z^*)^\kappa$ for some $\kappa$; in the case of stagnation points, $\kappa = 2$, while for corner singularities, $\kappa = \pi / \nu$, where $\nu$ is the in-fluid angle of the corner. In the analytic continuation of the free boundary, we have from the definition of the complex velocity, 
\begin{equation} \label{eq:gcwave_o2i_dwdz}
 \frac{dw}{dz} \sim q_0 e^{-i\theta_0} = q_0 = \text{const} \times (z-z^*)^{\kappa-1} =
\text{const} \times W^{(\kappa-1)/\kappa},
\end{equation}

\noindent and thus the inner limit of the outer solution is given by
\begin{equation} \label{eq:gcwave_o2i_q}
 q \sim c (w-w^*)^\alpha = cW^\alpha
\end{equation}

\noindent where $\alpha = (\kappa-1)/\kappa$ and $c$ is constant. 

%% Is zeta plane important?
%We will write $\zeta^*$ as the corresponding singularity in the $\zeta$-plane and then using the fact that $\zeta = e^{-w}$, we have for the case of a channel flow, $W \sim -(\zeta - \zeta^*)/\zeta^*$. 

Using $\chi$ in \eqref{eq:gcwave_chiprime} and the limiting form for $q_0$ in \eqref{eq:gcwave_o2i_q}, we see that $\chi$ must exhibit different limiting behaviours as the singularity is approached, depending on which sign of the square root is chosen and whether $\alpha$ is positive or negative:
\begin{equation}\label{eq:gcwave_o2i_chi}
\chi_\pm \sim \left\{ \begin{aligned} 
  \left[-\frac{ic}{\tau(\alpha+1)}\right] &W^{\alpha+1} 
 &&\equiv&& X_3^{\phantom{\pm}} W^{\alpha+1} 
 &&\text{for capillary $(+)$ and $\alpha < 0$} \\
	\left[-\frac{i}{c^3\beta(1-3\alpha)}\right] &W^{1-3\alpha} 
 &&\equiv&& X_2^{\phantom{\pm}}
 W^{1-3\alpha} &&\text{for gravity $(-)$ and $\alpha < 0$} \\
  \left[\mp \frac{i e^{\text{Arg}(\Delta)/2}}{c\sqrt{\beta\tau}(1-\alpha)}\right] &W^{1-\alpha}
 &&\equiv&& X_1^{\pm} W^{1-\alpha} &&\text{for $\alpha > 0$} \\
 \end{aligned} \right.
\end{equation}

For $\alpha < 0$, the choice of the negative sign for the square root leads to gravity waves, whereas the positive sign is associated with capillary waves; this is clear from the appearance of either $\beta$ (gravity) or $\tau$ (capillary) in the coefficients of $\chi$, but can also be understood by taking $\tau \to 0$, and noticing that the capillary root in \eqref{eq:gcwave_chiquadratic} disappears to infinity, leaving only the gravity root. 

There is a notable difficulty in determining the local behaviour of $\chi_\pm$ in \eqref{eq:gcwave_o2i_chi} when $\alpha > 0$, and this deals with the value of $\text{Arg}(\Delta)$, which must tend to $\pi$ or $-\pi$ as we approach the singularity from the upper half-$\zeta$-plane. For the purpose of the Stokes line analysis in the next section, it is sufficient to see that if we approach the singularity from along the upstream boundary, and set $W = |W|e^{i\pi}$, then $\Im(q_0^4 - A)$ tends to zero from below as $|W| \to 0$, and thus
\begin{equation} \label{argdelt}
e^{\text{Arg}(\Delta)/2} = -i.
\end{equation}

% For the moment, we shall leave the question of whether $e^{\text{Arg}(\Delta)} \sim 1$ or $\sim -1$, and determine the correct association (with either capillary or gravity waves) in the next section. 

% \noindent The constants $X_1, X_2$, and $X_3$ are defined for later use. Note that the $\mp$ sign in the first line is due to the fact that
% \[
% \pm \sqrt{\Delta} = \pm \sqrt{q_0^4 - A} \sim \mp i \sqrt{A},
% \]

% \noindent because $\text{Arg}(\Delta) = -\pi$, since the critical point is approached from the upper half-$\zeta$-plane. 

We also mention that the assumption that the potential follows a power of $(z - z^*)$ in \eqref{eq:gcwave_o2i_dwdz} will not be true for the case of a sink or source, where $W \sim \text{const.} \times \log(z - z^*)$. However, this analysis requires a somewhat different approach because the location of the singularity corresponds to $|w| \to \infty$. The problem of applying exponential asymptotics to gravity flow past a submerged line source was studied by \cite{lustri_2013}. 

\subsection{Local Stokes line analysis for the step} \label{sec:gcwave_stokes}

\noindent The existence of a Stokes line does not necessarily guarantee the appearance of surface waves. Indeed there are cases where the associated Stokes line fails to intersect the free-surface, as well as cases where the line lies on a different Riemann sheet altogether. In this section, we study the simplest scenario of a Stokes line which emerges from its singularity in the `in-fluid' direction of the upper-half $\zeta$-plane. Whether such a line actually encounters the free-surface ($\zeta\in\mathbb{R}^+$) is a global function of the leading-order flow and in general, must be checked by evaluating the integral in \eqref{eq:gcwave_chi}. For the moment, we ignore lines which leave in a direction of `into' the boundary ($\Im(\zeta) < 0$), and lines which enter secondary Riemann sheets. From a study of the global contours of $\chi$, we have observed that relevant interactions are generally governed by Stokes lines that traverse simple paths (from singularity to free-surface along the same Riemann sheet). However, in \S\ref{eq:gcwave_newsol} and Appendix \ref{sec:crossing}, we shall see the role played by more complicated Stokes line trajectories which cross onto different sheets.

Recall that the value of $\alpha$ in \eqref{eq:gcwave_o2i_q} determines the boundary's geometry near the singularity, with $\alpha < 0$ for a corner and $\alpha > 0$ for a stagnation point. The value of $c$ determines the orientation of the geometry, and this is not quite arbitrary because gravity provides a reference direction. Again, let us consider the two singularities associated with the step geometry in \eqref{eq:gcwave_rigidwall}.

For the case of the corner, with $\alpha \in (-1, 0)$ and $c \in\mathbb{R}^+$, then in the notation of \eqref{eq:gcwave_o2i_chi},
\begin{equation}
\chi_\pm \sim \left\{ \begin{aligned} 
\lvert X_3 \rvert & e^{-\pi i/2}W^{\alpha + 1} &\quad& \text{for capillary $(+)$} \\
 \lvert X_2 \rvert & e^{-\pi i/2}W^{1-3\alpha} &\quad& \text{for gravity $(-)$} 
 \end{aligned} \right.
\end{equation}

\noindent as $W \to 0$. If we write $\vartheta_\text{grav} = \text{Arg}(W)$ for the local angle of a gravity Stokes line, and $\vartheta_\text{cap}$ for the local angle of a capillary Stokes line, then we have from \eqref{eq:gcwave_chistokes} that
\[
\vartheta_\text{grav} = \pi \left(\frac{2m+1/2}{1-3\alpha}\right) \text{\qquad and \qquad}  
\vartheta_\text{cap} = \pi \left(\frac{2m+1/2}{\alpha+1} \right),
\]

\noindent where $m\in\mathbb{Z}$. As $\alpha$ increases from $-1$ to $0$, the first ($m = 0$) gravity Stokes line increases from $\vartheta_\text{grav} = \pi/8$ to $\pi/2$. However, for the same range in $\alpha$, the first capillary Stokes line only enters the upper-half plane beginning at $\alpha = -1/2$, where $\vartheta_\text{cap} = \pi$. As $\alpha$ increases to $0$, $\vartheta_\text{cap}$ increases to $\pi/2$. By taking the local analysis to next order in $W$, it can be shown in the case of the Stokes line initially leaving along the boundary, $\vartheta_\text{cap} = \pi$, it must continue along the boundary until it encounters the stagnation point. 

For the case of the stagnation point, with $\alpha \in (0, 1)$ and $c = |c|e^{-\pi i\alpha}$, then \eqref{eq:gcwave_o2i_chi} gives
\[
\chi_\pm \sim \mp \lvert X_1 \rvert e^{\pi i \alpha}W^{1-\alpha},
\]

\noindent as $W \to 0$ once we have used \eqref{argdelt}. The Stokes lines leave at angles of
\[
 \vartheta_\text{grav} = \pi \left(\frac{2m - \alpha}{1-\alpha}\right)
 \text{\quad \qquad and \qquad}
 \vartheta_\text{cap} = \pi \left(\frac{2m - 1 - \alpha}{1-\alpha}\right), 
\]

\noindent for $m\in\mathbb{Z}$. For gravity waves, there are no relevant Stokes
lines lying in the upper half-plane. For capillary waves, the only relevant
value of $m$ is $1$, which shows that there exists a Stokes line along $\theta =
\pi$ for all shapes. However, a secondary analysis to next order in $W$ shows
that the Stokes line which initially tends along $\vartheta_\text{cap} = \pi$ does indeed leave the boundary into the upper half-plane. 

The local Stokes line angles in the physical plane can be retrieved by using \eqref{eq:gcwave_o2i_dwdz} and \eqref{eq:gcwave_o2i_q}. If $z = z^*$ is the corresponding singularity, then 
\begin{equation} \label{eq:gcwave_z2w} 
z - z^* \sim \left[\frac{1}{c(1-\alpha)}\right] W^{1-\alpha}.
\end{equation}

In summary, given a step-up obstruction with initial inclination $\sigma\pi$, the following occurs: first, the stagnation point always produces a capillary Stokes line; second, the corner always produces a gravity Stokes line, and only produces a capillary Stokes line if $\sigma > 1/2$. Table \ref{tab:localstokes} provides an illustration of our local Stokes line analysis. Note that at $\alpha = 0$, there is no singularity, so the corresponding entry instead refers to the limiting process of $\alpha \to 0^-$. 

% \newcolumntype{Z}{>{\centering\arraybackslash}X}
% \renewcommand{\tabularxcolumn}[1]{>{\arraybackslash}m{#1}}
\begin{table} 
\begin{center} 
\includegraphics[width=1.0\textwidth]{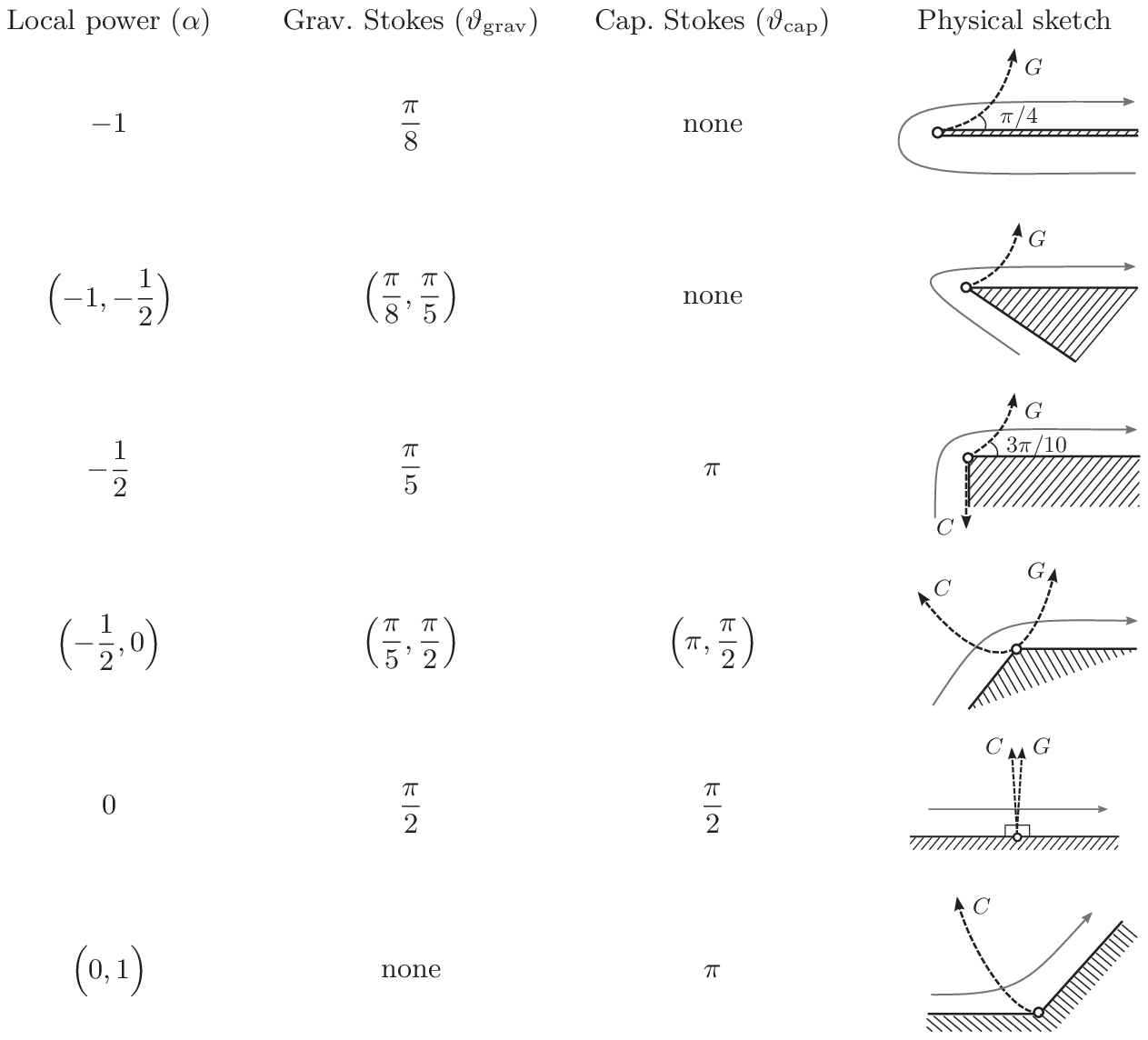}
 \end{center}
\caption{A summary of the local behaviour of a Stokes line (shown dashed) as it emerges from a singularity for which the complex velocity behaves like $dw/dz \sim c(w-w^*)^\alpha$. Angles $\vartheta$ are relative to the positive $\Re(w)$-axis. Angles in the physical plane follow from \eqref{eq:gcwave_z2w} and the solid arrow indicates the direction of flow.  \label{tab:localstokes}}
 \end{table} 

\section{Turning points and Stokes lines for the step} \label{sec:gcwave_tp}

\noindent The singularities encountered in \S\ref{sec:gcwave_chistokes} are not the only critical points in the gravity-capillary problem. From \eqref{eq:gcwave_DeltaA}, the \emph{turning points}, $\zeta = \zeta_T$, found at
\begin{equation} \label{eq:gcwave_DeltaA_repeat}
\Delta = q_0^4 - A = q_0^4 - \frac{4\tau}{\beta} = 0, 
\end{equation}

\noindent also represent a type of wave-generating singularity. These points do not share a connection with the physical geometry, but rather, they correspond to locations where derivatives of the capillary and gravity wavenumbers are equal. By drawing an analogy to the Airy equation \citep[Chap. 11]{white_book}, we would expect that these turning points can also produce Stokes lines, across which gravity and capillary waves can themselves interact. 

Care must be taken in solving $\Delta = 0$, however, since this generally involves the inversion of composite complex-valued functions with multiple branch cuts. In the case of the step flow \eqref{eq:gcwave_rigidwall}, for example, since $A > 0$, all the possible solutions are given by 
\begin{equation} \label{eq:gcwave_turningptinvert} 
\left(\frac{\zeta_T + b}{\zeta_T + a}\right) = A^{1/4\sigma} e^{2\pi i k/4\sigma}.
\end{equation}

\noindent Different values of $k \in \mathbb{Z}$ in \eqref{eq:gcwave_turningptinvert} may correspond to different turning points, but we must always check (\emph{a posteriori}) that the final location lies on the main Riemann sheet (\emph{i.e.} where $q_0$ is real and positive on $\zeta \in \mathbb{R}^+$). Like the previous Stokes line analysis of \S\ref{sec:gcwave_stokes}, we only consider points on this immediate sheet; numerical computations of the global contours of $\chi$ seem to indicate that critical points that lie on secondary sheets do not affect the free surface (though there are exceptions: see Appendix \ref{sec:crossing}). 

\begin{figure}
\includegraphics[width=1.0\textwidth]{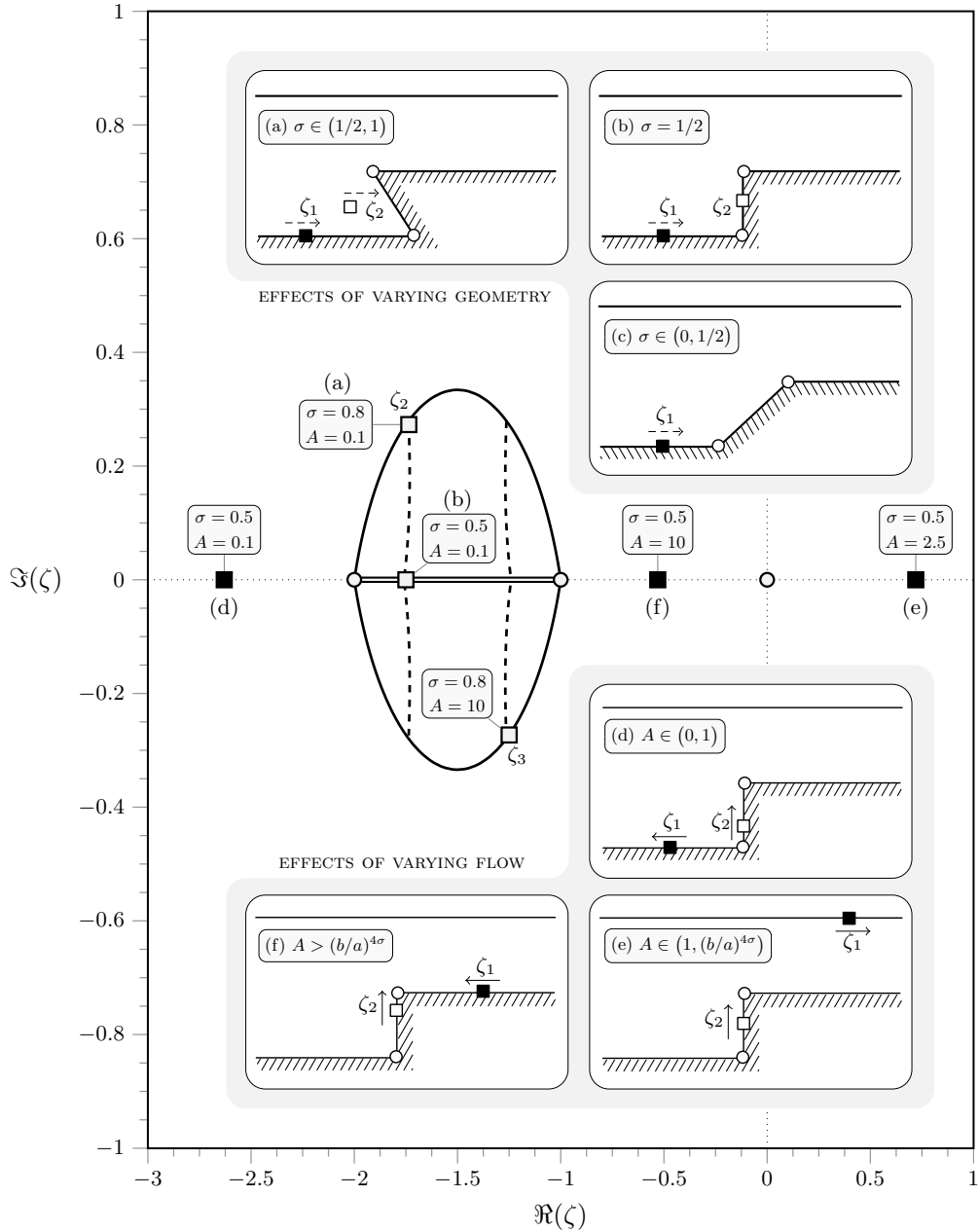}
 \caption{Movement of the turning points, $\zeta_1$, $\zeta_2$, $\zeta_3$ for the step with $b = 2$ (stagnation point) and $a = 1$ (corner). Exact values are shown in the $\zeta$-plane and the insets provide depictions in the physical plane. The $\zeta_1$ points are black squares, and $\zeta_{2, 3}$ are grey squares. For fixed values of $A$ and decreasing values of $\sigma$ (thus fixing the flow parameters, but changing the geometry), the different values of $\zeta_2$ and $\zeta_3$ are shown as a dashed line in the main illustration and through insets (a) to (c). Similarly, for changing values of $A$ and fixed values of $\sigma$ (thus changing the flow parameters, but fixing the geometry), the values of $\zeta_{2, 3}$ are shown as a solid line in the main illustration, and through insets (d)--(f). Note that for $\sigma < 0.5$, both $\zeta_{2}$ and $\zeta_3$ have passed onto the adjacent Riemann sheets, so they have disappeared from the physical illustration (c). \label{fig:gcwave_tp}}
\end{figure}

There are three important turning points in \eqref{eq:gcwave_turningptinvert}, which we denote as $\zeta_T = \zeta_1$, $\zeta_2$, and $\zeta_3$, corresponding to $k = 0$, $1$, and $-1$, respectively. Their locations for various values of $\sigma$ and $A$ are shown in the main portion of Figure \ref{fig:gcwave_tp}. Recall that the upper half-$\zeta$-plane shares a connection with the $z$-plane, and we have illustrated the physical connection within the insets (a) to (f). Note that we have taken the branch cut, marked by a striped line, along the real axis in the $\zeta$-plane, between the corner and stagnation points. 

First consider the two complex turning points, $\zeta_2$ and $\zeta_3$. For a fixed values of $A$ and $\sigma > 0.5$, the point $\zeta_2$ lies somewhere above the real axis, and $\zeta_3$ is its conjugate. For fixed $\sigma$ and as $A \to 0$, $\zeta_2$ tends to the stagnation point, while if $A \to \infty$, $\zeta_2$ tends to the corner point. This motion for variable $A$ is shown as a solid curve in Figure \ref{fig:gcwave_tp}. On the other hand, if $A$ is fixed and $\sigma \to 0.5$, $\zeta_2$ moves towards the branch cut, so that at $\sigma = 0.5$, $\zeta_2$ lies directly on the angled step, while $\zeta_3$ has moved onto the adjacent Riemann sheet. If $\sigma < 0.5$, then both $\zeta_1$ and $\zeta_3$ are now on secondary sheets. This motion for variable $\sigma$ is shown as the dashed line in the figure, and through the insets (a) to (c).

The turning point, $\zeta_T = \zeta_1$, is real for all values of $\sigma$ and $A$. When $A = 0$, $\zeta_1$ lies on top of the stagnation point, and as $A \to \infty$, $\zeta_1$ moves leftwards along the real $\zeta$-axis towards the corner point by passing through $|\zeta| = \infty$ and traversing the free-surface. There are two special points: $A = 1$ (when $\zeta_1$ has reached infinity upstream), and $A = (b/a)^{4\sigma}$ (when $\zeta_1$ has reached infinity downstream). This is illustrated for the case of a rectangular step in Figure \ref{fig:gcwave_tp}, and insets (d) to (f).

%In summary, there are two relevant turning points for $\sigma \in (0.5, 1)$ (one
%real, two complex), two relevant turning points for $\sigma = 0.5$ (both real),
%and only one turning point for $\sigma \in (0, 0.5)$ (real). 

\subsection{Inner analysis and reduction to the Airy Equation}
\label{sec:gcwave_tpinner}

Turning points are important for two reasons: first, a turning point which lies on the free surface, as in the case of $A > 1$, produces a change in one or both gravity and capillary waves from constant-amplitude oscillations to exponential decay [\emph{c.f.} $\Re(d\chi/dw$) in \eqref{eq:gcwave_chiprime}]; second, turning points also generate Stokes lines and can thus lead to the birth of new exponentials. Recall from \eqref{eq:gcwave_stokes_jumpq} that, upon crossing the \Mycirct{B} $>$ \Mycirct{G} or \Mycirct{B} $>$ \Mycirct{C} Stokes line from $\zeta = \zeta^*$, the base series switches on the exponential
\begin{equation} \label{eq:gcwave_tpqexp}
q_\text{exp} = -\frac{2\pi i Q_\pm(\zeta)}{\epsilon^\gamma} \exp\left[-
\frac{1}{\epsilon} \int_{\zeta^*}^{\zeta_T} \frac{d\chi_\pm}{d\zeta'} \ d\zeta' \right]
\exp\left[- \frac{1}{\epsilon}
\int_{\zeta_T}^{\zeta} \frac{d\chi_\pm}{d\zeta'} \ d\zeta' \right],
\end{equation}

\noindent which we have written so that the integration passes through an arbitrary turning point, $\zeta = \zeta_T$. The Stokes lines from the turning points are given at locations where one exponential (\emph{e.g.} gravity) reaches peak exponential dominance over the other (\emph{e.g.} capillary). Thus, 
\begin{equation}
\begin{alignedat}{3} \label{eq:gcwave_tpstokes} 
\text{\Mycirct{G}} > \text{\Mycirct{C}}: \qquad &
\Re\left[ - \int_{\zeta_T}^\zeta \frac{d\chi_-}{d\zeta'} \ d\zeta' \right] &&\geq 
\Re\left[ - \int_{\zeta_T}^\zeta \frac{d\chi_+}{d\zeta'} \ d\zeta' \right], \\
\text{\Mycirct{C}} > \text{\Mycirct{G}}: \qquad &
\Re\left[ - \int_{\zeta_T}^\zeta \frac{d\chi_+}{d\zeta'} \ d\zeta' \right] &&\geq 
\Re\left[ - \int_{\zeta_T}^\zeta \frac{d\chi_-}{d\zeta'} \ d\zeta' \right].
\end{alignedat}
\end{equation}

\noindent As $\zeta \to \zeta_T$, it is easy to verify from \eqref{eq:gcwave_turningptinvert} that the turning points are \emph{simple}, and so we may write 
\begin{equation} \label{eq:Deltalocal}
\Delta \sim D(\zeta-\zeta_T),
\end{equation}

\noindent where $D \in \mathbb{C}$. Also, from \eqref{eq:gcwave_Theta} and \eqref{eq:gcwave_Q}, we have that 
\begin{equation} \label{eq:gcwave_Q_turning}
Q_\pm \sim \frac{B_\pm}{(\zeta - \zeta_T)^{1/4}}
\end{equation}

\noindent where $B_\pm$ is constant. We now write \eqref{eq:gcwave_tpqexp} in a form that makes the connection to the Airy equation apparent. We introduce
\begin{equation} \label{eq:gcwave_G}
H_\pm(\zeta) \equiv -\frac{2\pi i}{\epsilon^\gamma} (\zeta - \zeta_T)^{1/4} Q_\pm \exp\left[- \frac{1}{\epsilon} \int_{\zeta^*}^{\zeta_T} \frac{d\chi_\pm}{d\zeta'} \ d\zeta' \right],
\end{equation}

\noindent so $H_\pm$ tends to a constant near the turning points. We also split the real and complex parts of $\chi$ in \eqref{eq:gcwave_chi} with $d\chi_\pm/d\zeta  = (S_1 \pm i S_2)$, where
 \begin{equation} \label{eq:gcwave_S1S2}
 S_1(\zeta) = - (dw/d\zeta)\frac{i q_0}{2\tau} \quad \ \text{and} \quad S_2(\zeta) = -(dw/d\zeta)\frac{\sqrt{\Delta}}{2\tau q_0}
 \end{equation}

\noindent and $dw/d\zeta = -1/\zeta$ is given by \eqref{wzeta} for the particular case of a channel geometry. The expression for $q_\text{exp}$ in \eqref{eq:gcwave_tpqexp} can now be written as 
\begin{equation} \label{eq:gcwave_qexp_rewrite}
q_\text{exp} =
\left[ H_\pm \exp\left(-\frac{1}{\epsilon} \int_{\zeta_T}^{\zeta} S_1 \ d\zeta' \right) \right] 
\times \frac{\exp\left(\mp \frac{i}{\epsilon} \int_{\zeta_T}^{\zeta} S_2 \ d\zeta' \right)}{(\zeta-\zeta_T)^{1/4}},
\end{equation}

\noindent Following the discussion of \S\ref{sec:gcwave_tp}, we shall assume that $\zeta = \zeta_T$ is a first-order turning point (although this may not be true of all possible geometries specified by $q_0$). If we like, we can re-scale $q$ and $\zeta$ near $\zeta = \zeta_T$ in order to remove the square-bracketed pre-factor of \eqref{eq:gcwave_qexp_rewrite}. The governing equations \eqref{eq:gcwave_bdint}--\eqref{eq:gcwave_dyn} would then reduce to an Airy equation (\emph{c.f.} \citealt{bender_book}), for which we can perform the local Stokes line analysis. However, it is faster to apply the following shortcut: using the local expression for $\Delta$ in \eqref{eq:Deltalocal}, and combining with \eqref{eq:gcwave_G}--\eqref{eq:gcwave_qexp_rewrite} gives
\begin{equation} \label{eq:gcwave_qexp_airy}
q_\text{exp} \sim 
H_\pm(\zeta_T) \, \frac{e^{\mp \frac{p}{\epsilon} (\zeta - \zeta_T)^{3/2}}}{(\zeta-\zeta_T)^{1/4}},
\end{equation}

\noindent valid in the limit $\zeta \to \zeta_T$, where $p\in\mathbb{C}$ is a constant that depends on $\zeta_T$. 
% \[
% p \equiv -\frac{i \sqrt{D}}{3\epsilon \tau q_0(\zeta_T)} \frac{dw}{d\zeta}\rvert_{\zeta = \zeta_T}.
% \]

We then note that the Airy switching requires
\begin{equation} \label{eq:gcwave_basicairy}
\frac{e^{\mp \frac{p}{\epsilon} (\zeta - \zeta_T)^{3/2}}}{(\zeta-\zeta_T)^{1/4}}
\Swtchsym{Turning Point}{C}{G} 
\frac{e^{\mp \frac{p}{\epsilon} (\zeta - \zeta_T)^{3/2}}}{(\zeta-\zeta_T)^{1/4}}
+ i M \frac{e^{\pm \frac{p}{\epsilon} (\zeta - \zeta_T)^{3/2}}}{(\zeta-\zeta_T)^{1/4}},
\end{equation}

\noindent that is, the dominant wave switches on the subdominant wave with a Stokes multiplier of $i$ \citep[Chap. 11]{white_book}. The quantity $M = \pm 1$ inserts a sign that depends on the direction of analytic continuation past the Stokes line (the standard Airy equation, for example, has $M = 1$ for analytic continuation in the counter-clockwise direction, relative to the turning point). 

In terms of the outer variables, $q$, this shows that as we analytically continue across a Stokes line from the turning point at $\zeta = \zeta_T$, the already existent capillary/gravity wave in \eqref{eq:gcwave_tpqexp}
\begin{equation} \label{eq:gcwave_qexp_again}
q_\text{exp} = \biggl[ -\frac{2\pi i Q_\pm}{\epsilon^\gamma} \biggr] \times 
 \exp\biggl[-\frac{1}{\epsilon} \int_{\zeta^*}^{\zeta} \frac{d\chi_\pm}{d\zeta'} \ d\zeta' \biggr],
\end{equation}

\noindent switches on a gravity/capillary wave, which satisfies the transition rule:
\begin{multline} \label{eq:gcwave_turningexp}
 q_\text{exp} \Swtchsym{Turning Point}{C}{G} 
q_\text{exp} + i M \biggl[-\frac{2\pi i Q_\mp}{\epsilon^\gamma}\biggr]
 \left[\frac{B_\pm}{B_\mp}\right] \\
\times \exp\biggl[-\frac{1}{\epsilon} \biggl( \int_{\zeta^*}^{\zeta_T} \frac{d\chi_\pm}{d\zeta'} \ d\zeta' + 
\int_{\zeta_T}^{\zeta^*} \frac{d\chi_\mp}{d\zeta'} \ d\zeta' \biggr) \biggr]  
\times \exp\left[-\frac{1}{\epsilon} \int_{\zeta^*}^{\zeta} \frac{d\chi_\mp}{d\zeta'} \ d\zeta' \right],
\end{multline}

\noindent where $B_\pm$ is given by \eqref{eq:gcwave_Q_turning}. We have written the integrals in \eqref{eq:gcwave_turningexp} so as to emphasise the turning-point process. Notice that the last exponential in \eqref{eq:gcwave_turningexp} clearly indicates the presence of a gravity/capillary wave [\eqref{eq:gcwave_qexp_again} with a negated exponential]. The first two integrals of \eqref{eq:gcwave_turningexp}, however, show that the original capillary/gravity wave was generated from $\zeta^*$ on the $\pm$ branch, went around the turning point $\zeta_T$, and returned to $\zeta^*$ along the $\mp$ branch. The values of most of the quantities (\emph{e.g.} $Q_\pm$, $B_\pm$, and $M$) will not be necessary for the following discussion, but can be numerically computed if required.

%\exp\left[-\frac{1}{\epsilon} \int_{\zeta^*}^{\zeta} \frac{d\chi_\pm}{d\zeta'} \ d\zeta' \right]

%\begin{multline}
% q_\text{exp} \sim -\frac{2\pi i Q_\pm (\zeta)}{\epsilon^\gamma} \exp\left[
%-\frac{1}{\epsilon} \int_{\zeta^*_{(0),(1)}}^{\zeta_{(0),(1)}} \frac{d\chi}{d\zeta'} \ d\zeta' \right]
%\\ \longmapsto 
%q_\text{exp} + 
% i \left[-\frac{2\pi i Q_\mp(\zeta)}{\epsilon^\gamma}\right]
% \left[\frac{B_\pm}{B_\mp}\right]
%\exp\left[-\frac{1}{\epsilon} \int_{\zeta^*_{(0),(1)}}^{\zeta_{(1),(0)}} \frac{d\chi}{d\zeta'} \ d\zeta' \right].
%\end{multline}
%
%\noindent The value in this expression is that it clearly shows that, for example, a capillary wave will switch on a gravity wave via an integration from the point $\zeta^*_{(0)}$ on the $k = 0^\text{th}$ Riemann sheet, rounding the turning point, and to the point $\zeta_{(1)}$ on the $k = 1^\text{st}$ Riemann sheet. 

%\[
% q_\text{exp} \sim \frac{2\pi i Q_\mp (\zeta)}{\epsilon^\gamma} \exp\left[
%-\frac{1}{\epsilon} \int_{\zeta^*}^{\zeta} \chi_\pm' d\zeta \right],
%\]

%\noindent switches on a new wave given by
%\begin{equation} \label{eq:gcwave_turningexp}
% i \left[\frac{2\pi Q_\pm(\zeta)}{\epsilon^\gamma}\right]
% \left[\frac{B_\pm}{B_\mp}\right]
%\exp\left[-\frac{1}{\epsilon} \int_{\zeta^*}^{\zeta_T} \chi_\pm' d\zeta \right] 
%\exp\left[-\frac{1}{\epsilon} \int_{\zeta_T}^{\zeta} \chi_\mp' d\zeta \right].
%\end{equation}

\section{New solutions for the rectangular step ($\sigma = 1/2$)} \label{eq:gcwave_newsol}

\noindent In the spirit of \cite{king_1987} and \cite{chapman_2006}, we now study the simplest case for a non-trivial step: the rectangular geometry. In Part 1 of our work, it was shown that for the case of a linearised step of small height, two solutions are possible at small Froude and Bond numbers: one with capillary waves upstream and gravity waves downstream, and the other with localised solitary waves with decaying oscillations in the far field. Furthermore, this bifurcation arises because the capillary and gravity Stokes lines coalesce. 

\begin{figure}
\includegraphics[width=1.0\textwidth]{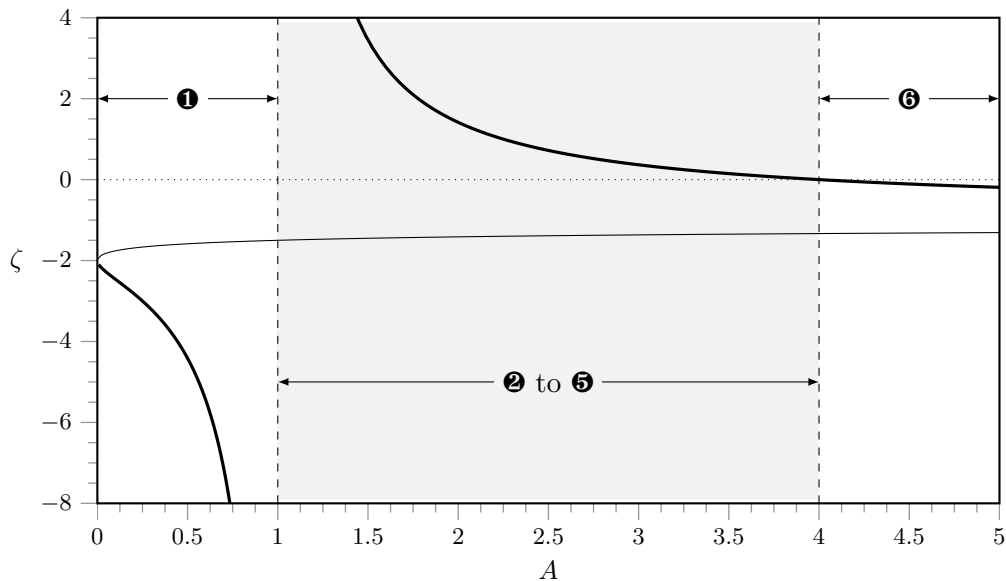}
\caption{Positions of the two turning points \eqref{zeta12} as a function of $A = 4\tau/\beta$ for the case of a rectangular step with $a = 1$ and $ b = 2$. The $T_1$ turning point at $\zeta = \zeta_1$ is shown thick, and the $T_2$ turning point at $\zeta = \zeta_2$ is shown thin. \label{fig:tprec}} 
\end{figure}

However, for the nonlinear step, there is the added subtlety in the form of the turning points. As we know from \S\ref{sec:gcwave_tp}, turning points not only change constant-amplitude waves to exponential decay, but also switch on secondary gravity or capillary waves. For the rectangular step, the two turning points are given by \eqref{eq:gcwave_turningptinvert}, simplified to
\begin{equation} \label{zeta12}
\zeta_1 = \frac{-b + a\sqrt{A}}{1-\sqrt{A}} \quad \text{and} \quad
\zeta_2 = \frac{-b - a\sqrt{A}}{1+\sqrt{A}},
\end{equation}

\noindent and these are shown in Figure \ref{fig:tprec} for the step with $a = 1$ and $b = 2$. Immediately, we recognise that there are at least three regions of interest: (i) both turning points on the solid boundary for $0 < A < 1$; (ii) one point on the surface and one on the solid boundary for $1 < A < (b/a)^2$; and both points back on the solid boundary for $A> (b/a)^2$.

In addition to these three regions, we must also note the positions of the Stokes lines relative to the turning points. It will be useful to introduce labels by
\begin{alignat*}{3}
C: \qquad & \text{\small Capillary Stokes line from stagnation point, $\zeta = -b$} \\
G: \qquad & \text{\small Gravity Stokes line from corner point, $\zeta = -a$} \\
\overline{T_1} \text{ and } \underline{T_1}: \qquad & \text{\small Capillary/gravity Stokes line from first turning point, $\zeta = \zeta_1$} \\
T_2: \qquad & \text{\small Capillary/gravity Stokes line from second turning point, $\zeta = \zeta_2$}
\end{alignat*}
%\begin{subequations}
% \begin{alignat*}{3}
% S: \qquad & \zeta = -b &\qquad & \text{\small (Stagnation point)} \\
% C: \qquad & \zeta = -a &\qquad & \text{\small (Corner point)} \\
% \overline{T_1} \text{ and } \underline{T_1}: \qquad & \zeta = \zeta_0 &\qquad & \text{\small (First turning point)} \\
% T_2: \qquad & \zeta = \zeta_1 &\qquad & \text{\small (Second turning point)}
% \end{alignat*}
%\end{subequations}

\noindent The underline, $\underline{T_1}$, is used to distinguish when the first turning point lies on bottom boundary and the overline, $\overline{T_1}$, for when it lies on the free-surface. 

\begin{figure}
\includegraphics[width=1.0\textwidth]{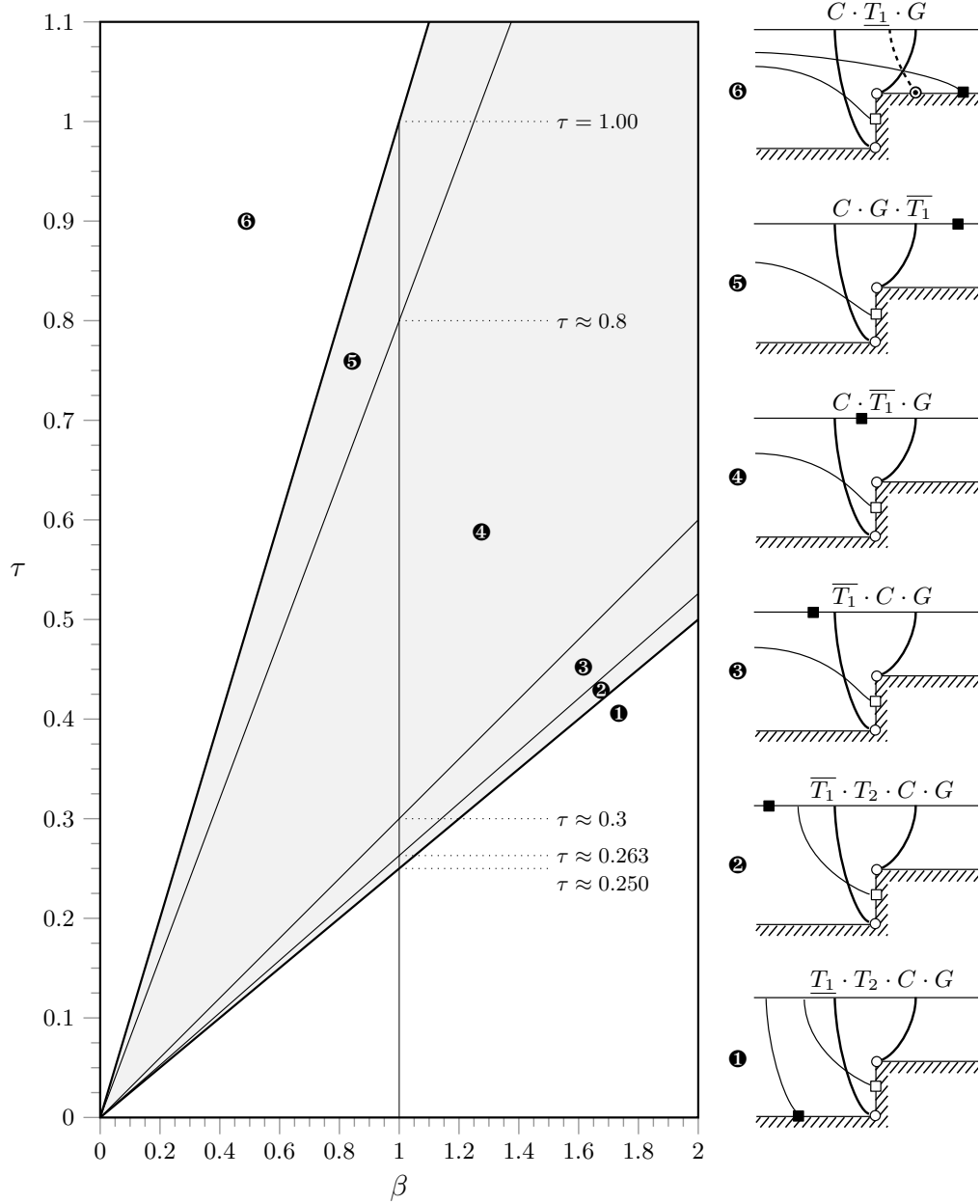}
\caption{The six regimes of interest for flow over a rectangular step, as shown in the $\beta\tau$-plane (left) for the case of $a = 1$ and $b = 2$. The gray strip is bounded by $A = 1$ and $A = (b/a)^2$, which thickens for larger and larger steps. The arrangement of Stokes lines and turning points is shown on the right. Stagnation and corner points are circular nodes and turning points are square nodes. The arrangement in Region 6 is complicated and will be discussed in \S\ref{sec:region6}. \label{fig:gcwave_betatau}}
\end{figure}

Next, examine the six subplots on the right of Figure \ref{fig:gcwave_betatau}, corresponding to six regions labelled \ding{182} and \ding{187}. These subplots indicate the qualitative arrangement of the turning points and Stokes lines for various values of $A > 0$. Solution-types are then classified by a sequence of labels, indicating the events as we traverse the free-surface from left to right (in the downstream direction). For example, a sequence like
\[
\underline{T_1} \cdot T_2 \cdot C \cdot G
\] 

\noindent in Region 1 corresponds to traversing the free-surface in the downstream direction, and crossing (i) the $\underline{T_1}$ Stokes line, (ii) the $T_2$ Stokes line, (iii) the capillary Stokes line from the stagnation point, and (iv) the gravity Stokes line
from the corner. 

Similarly, a sequence like
\[
C\cdot \overline{T_1} \cdot G
\]

\noindent in Region 4 corresponds to (i) crossing the $C$-line, (ii) crossing the $\overline{T_1}$ point on the free-surface, and then (iii) crossing the $G$-line. 

The sequence of subfigures in the right of Figure \ref{fig:gcwave_betatau} thus indicates that in the low-Froude, low-Bond limit, there are six possible solution types. Examine now the $\beta\tau$-plane shown in the left frame of Figure \ref{fig:gcwave_betatau}. The extremum cases of \ding{182} and \ding{187} correspond to the standard linearised solutions of Part 1 and confirms that the gray region of Figure \ref{fig:tprec}, which corresponds to $1 < A < (b/a)^2$, is further subdivided into \ding{183} to \ding{186}. Thus, this figure shows that the usual dispersion `line' which differentiates Rayleigh's two scenarios actually contains a \emph{series} of solutions \ding{183} to \ding{186}; the line's thickness and these new solutions are a manifestation of the nonlinearity (and largeness) of the step. As the step-height tends to zero, the line thickness tends to zero and we recover the standard linear picture. 

Remember that the physical sketches on the right subplots of Figure \ref{fig:gcwave_betatau} are only illustrative projections of the Stokes lines from the upper half-$\zeta$-plane onto the physical domain (see Figure 2 of Part 1). The reader should not forget that in crossing a Stokes line along the free surface, a wave contribution due to the singularities in the lower half-$\zeta$-plane also switch on; indeed the sum of these two contributions was used to derive \eqref{qexpthetaexp}. 

In the following discussion, it is sufficient to only consider the wave contributions from the upper half-$\zeta$-plane. We shorten the notation for the arguments of the exponentials in \eqref{eq:gcwave_stokes_jumpq} and \eqref{eq:gcwave_turningexp} by expressing the integrals relative to either the stagnation point, $\zeta = -a$, or the corner point, $\zeta = -b$. Thus
\begin{equation} \label{eq:gcwave_Xcapgrav}
X_\text{cap} = -\frac{1}{\epsilon} \int_{-a}^\zeta
\frac{d\chi_-}{d\zeta'} \, d\zeta' \text{\qquad and \qquad}
X_\text{grav} = -\frac{1}{\epsilon} \int_{-b}^\zeta 
\frac{d\chi_+}{d\zeta'} \, d\zeta'.
\end{equation}

\noindent We also denote the pre-factors for the exponentials by a calligraphic letter with a subscript, if applicable, which corresponds to an interaction with a turning point. This leads to the three symbols: 
\begin{align*}
\mathcal{A}&: \quad \text{An arbitrary wave amplitude to be determined} \\
\mathcal{C}&: \quad \text{A capillary wave amplitude generated from the stagnation point} \\
\mathcal{G}&: \quad \text{A gravity wave amplitude generated from the corner point}.
\end{align*}

Here are three examples: 
\[
\text{(i)} \ \mathcal{G} e^{X_\text{grav}} \qquad \quad
\text{(ii)} \ \mathcal{A}_{\overline{T_1} T_2} e^{X_\text{cap}} \qquad \quad
\text{(iii)} \ (\mathcal{A} + \mathcal{C})_{\overline{T_1}} e^{X_\text{grav}} 
\]

\noindent Example (i) denotes a gravity wave with amplitude $\mathcal{G}$ switched on after crossing the Stokes line from the corner. Example (ii) is constructed in a two-step process: an (arbitrary) capillary wave with amplitude $\mathcal{A}$ encounters a turning point and switches on a gravity wave with amplitude $\mathcal{A}_{\overline{T_1}}$, then encounters the second turning point and switches on a capillary wave with amplitude $\mathcal{A}_{\overline{T_1} T_2}$. Example (iii) is constructed by beginning with a capillary wave with combined pre-factors $\mathcal{A}$ and $\mathcal{C}$ (the former an arbitrary quantity, and the latter generated from the stagnation point), and then switching on a gravity wave due to the $\overline{T_1}$ turning point. Our notation and the associated expressions will be made clear in the many examples and figures to follow.

% \noindent We also denote the pre-factors for the exponentials by a calligraphic letter with a subscript for the associated singularity, and, if applicable, a superscript for the associated turning point(s). Pre-factors not accompanied with a subscript or superscript correspond to a general wave, whose form we shall eventually derive. Here are three examples: 
% \[
% \text{(i)} \ \mathcal{G} e^{X_\text{grav}} \qquad \quad
% \text{(ii)} \ \mathcal{A}_{\overline{T_1} T_2} e^{X_\text{cap}} \qquad \quad
% \text{(iii)} \ (\mathcal{A} + \mathcal{C})^{\overline{T_1}} e^{X_\text{grav}} 
% \]

% \noindent Example (i) denotes a gravity wave with amplitude $\mathcal{G}$ switched-on after crossing the Stokes line from the corner. Example (ii) is constructed in a two-step process: an (arbitrary) capillary wave with amplitude $\mathcal{A}$ encounters a turning point and switches-on a gravity wave with amplitude $\mathcal{A}_{\overline{T_1}}$, then finally encounters the second turning point and switches-on a capillary wave with amplitude $\mathcal{A}_{\overline{T_1} T_2}$. Example (iii) is constructed by beginning with a capillary wave with combined pre-factors $\mathcal{A}$ and $\mathcal{B_S}$ (the former an arbitrary quantity, and the latter generated from the stagnation point), and then switching-on a gravity wave due to the $\overline{T_1}$ turning point. Our notation and the associated expressions will be made clear in the many examples and figures to follow.

\subsection{Region 1 with $\underline{T_1}\cdot T_2\cdot C\cdot G$}
%\subsection{Region {\ding{182}}}
\label{sec:gcwave_region1}
 
\noindent Consider the case of $A < 1$. For most (larger) values of $\tau$, as we analytically continue across the free-surface, the sequence of events is $\underline{T_1}\cdot T_2\cdot C\cdot G$, which involves the $T_2$ Stokes line crossing the $C$ Stokes line. In fact, for extremely small values of $\tau$, the lines may not actually cross, and so the sequence is $\underline{T_1} \cdot C\cdot T_2\cdot G$; however, both cases are equivalent because, as we shall see, the turning points do not play any role in determining the free surface. 

In order to derive the correct wave expressions, we traverse the free-surface, from left-to-right, beginning with a wave that satisfies the upstream radiation condition (\emph{i.e.} a capillary wave). Upon reaching the end, we apply the downstream radiation condition, and this process provides us with a complete set of fully-determined waves. If we start with an arbitrary upstream capillary wave $\mathcal{A}e^{X_\text{cap}}$, then the sequence of events is
\begin{eqnarray*}
\mathcal{A} e^{X_\text{cap}}  
&\Swtch{$\underline{T_1}$}{G}{C} & 
\mathcal{A} e^{X_\text{cap}} \\
&\Swtch{$T_2$}{G}{C} & 
\mathcal{A} e^{X_\text{cap}} \\
&\Swtch{Stag.}{B}{C}&
(\mathcal{A}+\mathcal{C})e^{X_\text{cap}} \\
&\Swtch{Corn.}{B}{G} &
(\mathcal{A}+\mathcal{C})e^{X_\text{cap}} +
\mathcal{G} e^{X_\text{grav}}.
\end{eqnarray*}

\begin{figure}
\includegraphics[width=1.0\linewidth]{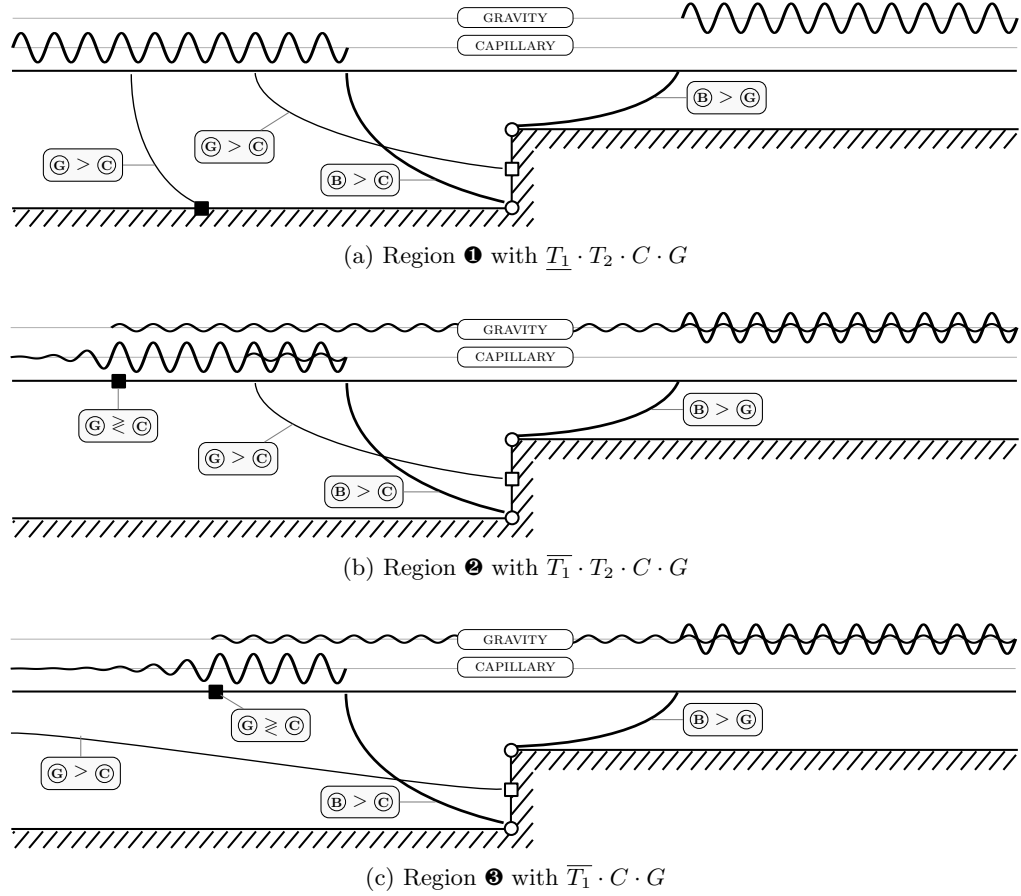}
\caption{Sketch of gravity-capillary solutions in Regions \ding{182} to \ding{184}. Waves sketched with the larger amplitudes are switched on by the stagnation point or corner; waves sketched with smaller amplitudes are switched on by turning points. Note that the gravity and capillary waves differ in both amplitude and wavelength, but we have not illustrated this difference.\label{fig:gcwaves_region123}}
%The waves in Region \ding{182} are identifiable with the first of Rayleigh's two linear solutions (valid for small steps), with constant-amplitude capillary waves upstream and constant-amplitude gravity waves downstream.}  
\end{figure}

\noindent There are no capillary waves downstream, so $\mathcal{A} +
\mathcal{C} = 0$, and the final solution has
\begin{equation*}
-\mathcal{C} e^{X_\text{cap}} \Swtch{Stag.}{B}{C} 0
\Swtch{Corn.}{B}{G} \mathcal{G}e^{X_\text{grav}},
\end{equation*}

\noindent where the pre-factors, $\mathcal{C}$ and $\mathcal{G}$, are given by the jump condition \eqref{eq:gcwave_stokes_jumpq}, with
\begin{equation} \label{eq:gcwave_region1_AC}
\mathcal{C} = -\frac{2\pi i Q_+}{\epsilon^{\gamma_C}}
\text{\quad and \quad}
\mathcal{G} = -\frac{2\pi i Q_-}{\epsilon^{\gamma_G}}
\end{equation}

\noindent and $Q_\pm$ is computed from \eqref{eq:gcwave_Theta} and \eqref{eq:gcwave_Q}. For the rectangular step, it can be shown by matching inner and outer solutions that $\gamma_C = 0$ and $\gamma_G = 6/5$. In the end, the free-surface resembles the standard linearised solution with capillary waves upstream and gravity waves downstream, but now, with a wave-free region near the centre, similar to what occurs in Part 1 of our work. The result is sketched in Figure \ref{fig:gcwaves_region123}(a). Note that the Stokes lines from the turning points remain inactive.

In some problems, the intersection of Stokes lines can lead to more subtle effects, such as the higher-order Stokes Phenomenon (see for example, \citealt{howls_2004}). A simple consistency check is to analytically continue in a closed path around the intersection point, and to see if the solution returns to its original value. In this case, we see that the $T_2$ Stokes line remains inactive, and so the intersection of the \Mycirct{B} $>$ \Mycirct{C} and \Mycirct{G} $>$ \Mycirct{C} lines are not a concern. This will not be the case for the solution in Region 6.

\subsection{Region 2 with $\overline{T_1}\cdot T_2\cdot C\cdot G$}
%\subsection{Region {\ding{183}}}

\noindent As $\tau$ is increased and $\beta$ held steady, the $\underline{T_1}$ turning point moves upstream along the solid boundary and reaches $-\infty$ at $A = 1$. Solutions in Region 2 have $A > 1$ and the turning point $\overline{T_1}$ has now moved onto the free-surface and travels downstream as $\tau$ is further increased. The turning point on the free-surface also implies that upstream waves will now decay, but downstream waves will remain constant. We start with an arbitrary upstream capillary wave $\mathcal{A} e^{X_\text{cap}}$, and the sequence of events is:
\begin{eqnarray*}
\mathcal{A} e^{X_\text{cap}}  
&\Swtchsym{$\overline{T_1}$}{G}{C}& 
\mathcal{A} e^{X_\text{cap}} + \mathcal{A}_{\overline{T_1}} e^{X_\text{grav}} \\
&\Swtch{$T_2$}{G}{C}& 
(\mathcal{A} + \mathcal{A}_{\overline{T_1} T_2}) e^{X_\text{cap}} + \mathcal{A}_{\overline{T_1}}
e^{X_\text{grav}} \\
&\Swtch{Stag.}{B}{C}& 
(\mathcal{A} + \mathcal{A}_{\overline{T_1} T_2} + \mathcal{C}) e^{X_\text{cap}} +
\mathcal{A}_{\overline{T_1}} e^{X_\text{grav}} \\
&\Swtch{Corn.}{B}{G}& 
(\mathcal{A} + \mathcal{A}_{\overline{T_1} T_2} + \mathcal{C}) e^{X_\text{cap}} +
(\mathcal{A}_{\overline{T_1}}+\mathcal{G}) e^{X_\text{grav}}.
\end{eqnarray*}

%\noindent In the first line, the $\overline{T_1}$ transition follows directly from \eqref{eq:gcwave_turningexp}, but in the second line, the $T_2$ transition has a negated  ($- i \cdot i$) coefficient---this is because crossing the Stokes line in the downstream direction is analogous to analytically continuing the Airy functions in the clockwise direction. 

\noindent The last transition is known by \eqref{eq:gcwave_stokes_jumpq}, so $\mathcal{G}$ is again given by \eqref{eq:gcwave_region1_AC}. Since there are no capillary waves downstream, and $\mathcal{C}$ is provided by the jump condition in \eqref{eq:gcwave_stokes_jumpq}, then
\begin{equation} \label{eq:gcwave_region2_Bs}
\mathcal{C} = -\mathcal{A}  - \mathcal{A}_{\overline{T_1} T_2} = -\frac{2\pi i Q_+}{\epsilon^{\gamma_C}}. 
\end{equation}

\noindent We can associate the pre-factor $\mathcal{A}$ with the stagnation point, writing
\begin{equation} \label{eq:gcwave_region2_Abar}
\mathcal{A}(\zeta) = \left[- \frac{2\pi i Q_+}{\epsilon^{\gamma_C}} \right] \widehat{\mathcal{A}}(\zeta),
\end{equation}

\noindent for some unknown function, $\widehat{A}(\zeta)$. The pre-factor, $\mathcal{A}_{\overline{T_1}}$, of the first gravity wave switched on by the $\overline{T}_1$ turning-point is given by the transition rule \eqref{eq:gcwave_turningexp}, and this allows us to relate $\widehat{\mathcal{A}}$ to $\mathcal{A}_{\overline{T_1}}$. Care must be taken to express the gravitational wave quantities with respect to the corner, instead of the stagnation point [see the definition of $X_\text{grav}$ in \eqref{eq:gcwave_Xcapgrav}]. Next, the pre-factor, $\mathcal{A}_{\overline{T_1}T_2}$, of the second capillary wave switched on by the $T_2$ turning point is also given by \eqref{eq:gcwave_turningexp}, which relates $\widehat{\mathcal{A}}$ to $\mathcal{A}_{\overline{T_1}T_2}$. Lastly, \eqref{eq:gcwave_region2_Bs} and \eqref{eq:gcwave_region2_Abar} allows $\mathcal{A}$ to be solved entirely in terms of $Q_\pm$ and thus all quantities can be determined.

Therefore, far upstream we have a decaying capillary wave,
while far downstream we have a constant gravity wave. Also, switched on after
encountering the $\overline{T_1}$ turning point is a (doubly) exponentially
small gravity wave, which continues downstream. Finally, there is an even
smaller capillary wave which has been turned on by the gravity wave
across the $T_2$-line. This is illustrated in Figure \ref{fig:gcwaves_region123}(b), and in summary,
\begin{multline}
\mathcal{A} e^{X_\text{cap}}  
\Swtchsym{$\overline{T_1}$}{G}{C}
\mathcal{A} e^{X_\text{cap}} + \mathcal{A}_{\overline{T_1}} e^{X_\text{grav}}
\Swtch{$T_2$}{G}{C}
(\mathcal{A} + \mathcal{A}_{\overline{T_1}T_2}) e^{X_\text{cap}} + \mathcal{A}_{\overline{T_1}}
e^{X_\text{grav}} \\
\Swtch{Stag.}{B}{C}
\mathcal{A}_{\overline{T_1}} e^{X_\text{grav}}
\Swtch{Corn.}{B}{G}
(\mathcal{A}_{\overline{T_1}}+\mathcal{G}) e^{X_\text{grav}}. 
\end{multline}

%\begin{figure}
%%\includegraphics[width=1.0\linewidth]{figpdf/gcwave_region2}
%\begin{tikzpicture}
%\node[inner sep=0pt, outer sep=0pt] at (0,0) {\includegraphics[width=1.0\linewidth]{figpdf/gcsketch2.pdf}};
%
%  \node[style=benode] at (0,1.27) {\scshape gravity};
%  \node[style=benode] at (0,0.9) {\scshape capillary};
%
%%\node[scale=0.7] at (-6.2,1.3) {\scshape gravity};
%%\node[scale=0.7] at (6.1,0.80) {\scshape capillary};
%%\draw[very thin] (-5.7,1.30) -- (-5.45,1.23);
%%\draw[very thin] (5.45,0.80) -- (5.15,0.90);
%   \node[pin={[pin distance=10pt] 270:{\mycircd{-6pt}{G} $\gtrless$ \mycircd{-5.7pt}{C} }}] at (-5.3, 0.6) {};
%   \node[pin={[pin distance=10pt] 225:{\mycircd{-5pt}{G} $>$ \mycircd{-5pt}{C} }}] at (-2.85, 0.08) {};
%   \node[pin={[pin distance=17pt] 180:{\mycircd{-2pt}{B} $>$ \mycircd{-2pt}{C} }}] at (-0.9, -0.9) {};
%   \node[pin={[pin distance=15pt] 0:{\mycircd{-2pt}{B} $>$ \mycircd{-2pt}{G} }}] at (1.7, 0.2) {};
%\end{tikzpicture}
%\captionof{figure}{Region \ding{183} with $\overline{T_1}\cdot T_2\cdot C\cdot G$. Once the turning point has passed upstream, the capillary waves decay in the far field. The two turning points are responsible for producing a second pair of gravity and capillary waves. \label{fig:gcwaves_region2}}
%\end{figure}

In contrast to Region 1, the $T_2$ line is now active, but the intersection of the two Stokes lines remains unproblematic; we can still analytically continue in a closed path around the intersection point and return to our original value. 

We note that the solutions in this region are actually contained in a rather small section of $\beta\tau$-space (as shown in Figure \ref{fig:gcwave_betatau}). Once $\overline{T_1}$ has crossed the point that marks the intersection of the $T_2$ line and the free-surface, a bifurcation occurs and the $T_2$ line tends to $w = -\infty$ without intersecting the free-surface; this can be verified by numerical integration.

%The reason for this is that near $A = 1$, the location of the $\overline{T_1}$ turning point is 
%\[
%w \sim -\log \biggl[\frac{2(b-a)}{A-1}\biggr], 
%\]

%\noindent and so it quickly increases from $w = -\infty$ to larger values. 

\subsection{Region 3 with $\overline{T_1}\cdot C\cdot G$}
%\subsection{Region {\ding{184}}}

In Region 3, the only difference to the previous region is that the $T_2$-line tends to $w = -\infty$ without intersecting the free surface. Starting with an arbitrary upstream capillary wave, $\mathcal{A} e^{X_\text{cap}}$, the sequence of events is now:
\begin{eqnarray*}
\mathcal{A} e^{X_\text{cap}}  
&\Swtchsym{$\overline{T_1}$}{G}{C}
& \mathcal{A} e^{X_\text{cap}} +
\mathcal{A}_{\overline{T_1}} e^{X_\text{grav}} \\
& \Swtch{Stag.}{B}{C}
& (\mathcal{A} + \mathcal{C}) e^{X_\text{cap}} + \mathcal{A}_{\overline{T_1}} e^{X_\text{grav}} \\
& \Swtch{Corn.}{B}{G} & 
(\mathcal{A} + \mathcal{C})
e^{X_\text{cap}} + (\mathcal{A}_{\overline{T_1}}+\mathcal{G}) e^{X_\text{grav}}.
\end{eqnarray*}

\noindent The downstream radiation condition requires $\mathcal{C} = -\mathcal{A}$. Like the previous regions, \eqref{eq:gcwave_region1_AC} provides the expressions for $\mathcal{C}$ and $\mathcal{G}$, and \eqref{eq:gcwave_turningexp} relates $\mathcal{A}$ to $\mathcal{A}_{\overline{T_1}}$. The final result is
\begin{multline*}
\qquad \qquad \mathcal{A} e^{X_\text{cap}} 
\Swtchsym{$\overline{T_1}$}{G}{C}
\mathcal{A} e^{X_\text{cap}} + \mathcal{A}_{\overline{T_1}} e^{X_\text{grav}} \\
\Swtch{Stag.}{B}{C}
\mathcal{A}_{\overline{T_1}} e^{X_\text{grav}}
\Swtch{Corn.}{B}{G}
(\mathcal{A}_{\overline{T_1}}+\mathcal{G}) e^{X_\text{grav}}. \qquad \qquad
\end{multline*}

\noindent Thus the only difference between the solutions in Regions 2 and 3 is that in the latter, the doubly-exponentially-small capillary wave has disappeared from the free surface. A sketch of the solution is shown in Figure \ref{fig:gcwaves_region123}(c).

%\begin{figure}
%\includegraphics[width=1.0\textwidth]{finalfig/gcwaves_region3.pdf}
%\caption{}
%\end{figure}

%\begin{figure}
%%\includegraphics[width=1.0\linewidth]{figpdf/gcwave_region3}
%\begin{tikzpicture}
%\node[inner sep=0pt, outer sep=0pt] at (0,0) {\includegraphics[width=1.0\linewidth]{figpdf/gcsketch3.pdf}};
%
%  \node[style=benode] at (0,1.27) {\scshape gravity};
%  \node[style=benode] at (0,0.9) {\scshape capillary};
%
%   %\draw[very thin] (-5.7,1.33) -- (-5.45,1.23);
%   %\draw[very thin] (5.45,0.85) -- (5.20,0.95);
%
%   \node[pin={[pin distance=7pt] 315:{\mycircd{-6pt}{G} $\gtrless$ \mycircd{-5.7pt}{C} }}] at (-4.05, 0.6) {};
%   \node[pin={[pin distance=10pt] 270:{\mycircd{-6pt}{G} $>$ \mycircd{-5.7pt}{C} }}] at (-5.8, 0.1) {};
%   \node[pin={[pin distance=17pt] 180:{\mycircd{-2pt}{B} $>$ \mycircd{-2pt}{C} }}] at (-0.95, -0.9) {};
%   \node[pin={[pin distance=15pt] 0:{\mycircd{-2pt}{B} $>$ \mycircd{-2pt}{G} }}] at (1.7, 0.2) {};
%\end{tikzpicture}
%\captionof{figure}{Region \ding{184} with $\overline{T_1}\cdot C\cdot G$. This region is distinguished from the previous by virtue of the $T_2$-line, which no longer intersects the free surface. Apart from the disappearance of a subdominant capillary wave, the solutions in this region are the same as the previous.  \label{fig:gcwaves_region3}}
%\end{figure}

\subsection{Region 4 with $C\cdot \overline{T_1}\cdot G$}
%\subsection{Region {\ding{185}}}

\noindent For $A$ still larger, $\overline{T_1}$ eventually crosses the intersection point of the $C$-line and the free-surface, and then the sequence is then $C\cdot\overline{T_1}\cdot G$. If we start with an arbitrary upstream capillary wave $\mathcal{A} e^{X_\text{cap}}$, the sequence of events is
\begin{eqnarray*}
\mathcal{A} e^{X_\text{cap}}  
& \Swtch{Stag.}{B}{C}
& (\mathcal{A} + \mathcal{C}) e^{X_\text{cap}} \\
&\Swtchsym{$\overline{T_1}$}{G}{C}
& (\mathcal{A} + \mathcal{C}) e^{X_\text{cap}} +
(\mathcal{A}+\mathcal{C})_{\overline{T_1}} e^{X_\text{grav}} \\
& \Swtch{Corn.}{B}{G}
& (\mathcal{A} + \mathcal{C}) e^{X_\text{cap}} +
((\mathcal{A}+\mathcal{C})_{\overline{T_1}} + \mathcal{G}) e^{X_\text{grav}}.
\end{eqnarray*}

\noindent Again, we cannot have capillary waves downstream, so $\mathcal{A} + \mathcal{C} = 0$, and the coefficients, $\mathcal{C}$ and $\mathcal{G}$, are given by \eqref{eq:gcwave_region1_AC}. The final result is
\[
-\mathcal{C} e^{X_\text{cap}}
\Swtch{Stag.}{B}{C} 0 
\Swtch{Corn.}{B}{G} \mathcal{G} e^{X_\text{grav}}.
\]

\begin{figure}
\includegraphics[width=1.0\linewidth]{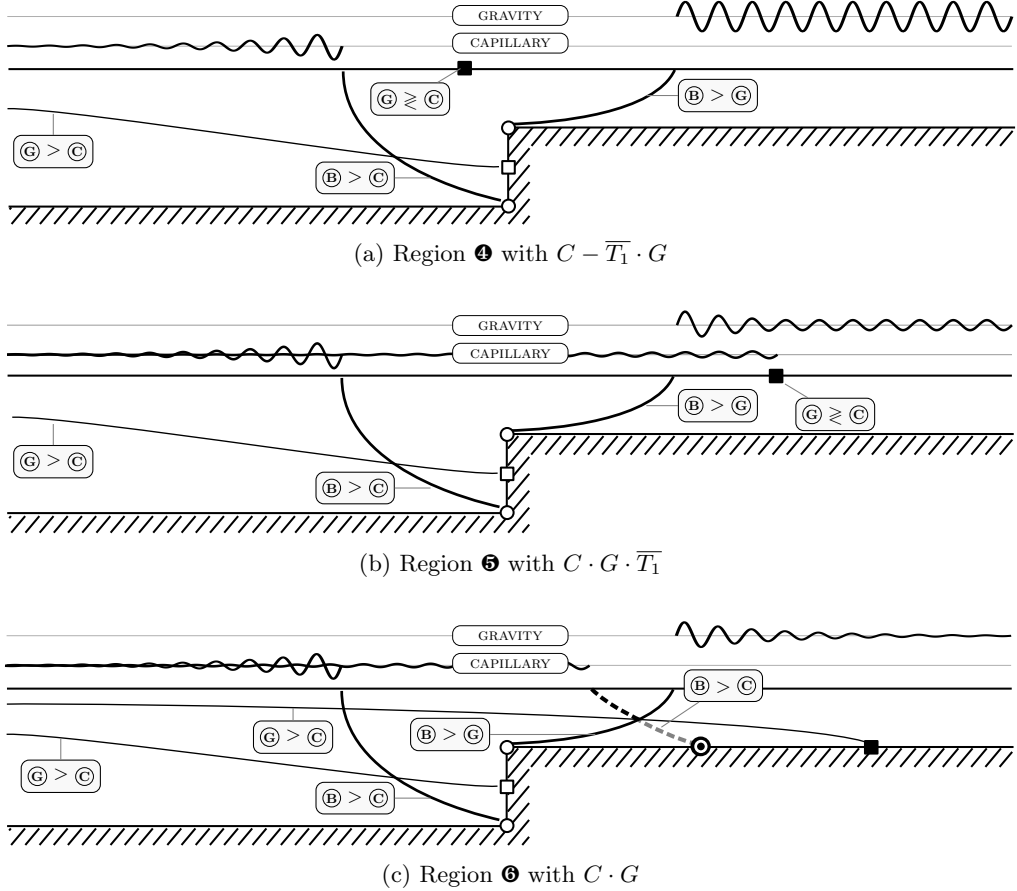}

\caption{Sketch of gravity-capillary solutions in Regions \ding{185} to \ding{187}. The solution in Region \ding{187} should correspond to the second of Rayleigh's linearised profiles, but is complicated by the two crossing Stokes lines. In Appendix \ref{sec:crossing}, we demonstrate the existence of a third Stokes line (dashed), which originates from a secondary singularity (black circle). \label{fig:gcwaves_region456}} 
\end{figure}

\noindent The solution is sketched in Figure \ref{fig:gcwaves_region456}(a), and we see that it consists of a decaying capillary wave upstream and a constant amplitude gravity wave downstream. 

%\begin{figure}
%%\includegraphics[width=1.0\linewidth]{figpdf/gcwave_region4}
%\begin{tikzpicture}
%\node[inner sep=0pt, outer sep=0pt] at (0,0) {\includegraphics[width=1.0\linewidth]{figpdf/gcsketch4.pdf}};
%  
%  \node[style=benode] at (0,1.27) {\scshape gravity};
%  \node[style=benode] at (0,0.9) {\scshape capillary};
%  
%   \node[pin={[pin distance=7pt] 225:{\mycircd{-6pt}{G} $\gtrless$ \mycircd{-5.7pt}{C} }}] at (-0.55, 0.63) {};
%   \node[pin={[pin distance=10pt] 270:{\mycircd{-6pt}{G} $>$ \mycircd{-5.7pt}{C} }}] at (-6.1, 0.07) {};
%   \node[pin={[pin distance=17pt] 180:{\mycircd{-2pt}{B} $>$ \mycircd{-2pt}{C} }}] at (-0.95, -0.9) {};
%   \node[pin={[pin distance=15pt] 0:{\mycircd{-2pt}{B} $>$ \mycircd{-2pt}{G} }}] at (1.7, 0.2) {};
%\end{tikzpicture}
%\captionof{figure}{Region \ding{185} with $C-\overline{T_1}\cdot G$. Once the turning point has passed the Stokes line from the stagnation point, solutions simply contain a decaying capillary wave upstream and a constant amplitude gravity wave downstream. \label{fig:gcwaves_region4}}
%\end{figure}

\subsection{Region 5 with $C\cdot G\cdot \overline{T_1}$}

\noindent Eventually, $\overline{T_1}$ passes the point where the $G$-line intersects the free-surface, and so downstream from this point, the gravity waves switch from constant-amplitude to exponentially decaying oscillations. For solutions in this region, it is easier to analytically continue from downstream to upstream, and negate each of the transitions [since the rules \eqref{eq:gcwave_stokes_jumpq} and \eqref{eq:gcwave_turningexp} are written for continuation in the downstream direction]. If we start with an arbitrary downstream gravity wave $\mathcal{A} e^{X_\text{grav}}$, the
sequence follows
\begin{eqnarray*}
\mathcal{A} e^{X_\text{grav}}  
& \Swtchsymback{$\overline{T_1}$}{G}{C}
& -\mathcal{A}_{\overline{T_1}} e^{X_\text{cap}} + \mathcal{A} e^{X_\text{grav}}  \\
& \Swtchback{Corn.}{B}{G}
& -\mathcal{A}_{\overline{T_1}} e^{X_\text{cap}} + (\mathcal{A} -\mathcal{G})e^{X_\text{grav}}  \\
& \Swtchback{Stag.}{B}{C}
& (-\mathcal{A}_{\overline{T_1}} - \mathcal{C})e^{X_\text{cap}} + (\mathcal{A}
-\mathcal{G})e^{X_\text{grav}}.
\end{eqnarray*}

\noindent We impose the requirement that there are no gravity waves downstream, so $\mathcal{A} - \mathcal{G} = 0$. The values $\mathcal{B_S}$ and $\mathcal{C_C}$ are given in \eqref{eq:gcwave_region1_AC}, and $\mathcal{A}$ and $\mathcal{A}_{\overline{T_1}}$ are related through  \eqref{eq:gcwave_turningexp}. The final result is given by (from left to right)
\[
(-\mathcal{A}_{\overline{T_1}} - \mathcal{C})e^{X_\text{cap}}  
\Swtch{Stag.}{B}{C}
-\mathcal{A}_{\overline{T_1}}e^{X_\text{cap}} 
\Swtch{Corn.}{B}{G}
-\mathcal{A}_{\overline{T_1}} e^{X_\text{cap}} + \mathcal{A} e^{X_\text{grav}}
\Swtchsym{$\overline{T_1}$}{G}{C}
\mathcal{A} e^{X_\text{grav}}. 
\]

\noindent The solution is shown in Figure \ref{fig:gcwaves_region456}(b), where we see it consists of two decaying capillary waves upstream, and a gravity wave downstream. The gravity wave, which is first switched on by the Stokes line from the corner, decays until it reaches the turning point, and then switches to a constant amplitude wave downstream. 

%\begin{figure}
%%\includegraphics[width=1.0\linewidth]{figpdf/gcwave_region5}
%\begin{tikzpicture}
%\node[inner sep=0pt, outer sep=0pt] at (0,0) {\includegraphics[width=1.0\linewidth]{figpdf/gcsketch5.pdf}};
%
%  \node[style=benode] at (0,1.27) {\scshape gravity};
%  \node[style=benode] at (0,0.9) {\scshape capillary};
%
%   \node[pin={[pin distance=7pt] 315:{\mycircd{-6pt}{G} $\gtrless$ \mycircd{-5.7pt}{C} }}] at (3.55, 0.56) {};
%   \node[pin={[pin distance=10pt] 270:{\mycircd{-6pt}{G} $>$ \mycircd{-5.7pt}{C} }}] at (-6.1, 0.07) {};
%   \node[pin={[pin distance=17pt] 180:{\mycircd{-2pt}{B} $>$ \mycircd{-2pt}{C} }}] at (-0.95, -0.9) {};
%   \node[pin={[pin distance=15pt] 0:{\mycircd{-2pt}{B} $>$ \mycircd{-2pt}{G} }}] at (1.7, 0.2) {};
%\end{tikzpicture}
%\captionof{figure}{Region \ding{186} with $C\cdot G\cdot \overline{T_1}$. The turning point has now passed the $G$-Stokes line, and causes a small segment of the gravity waves to decay. However, further downstream from the turning point, the gravity waves are of constant amplitude. \label{fig:gcwaves_region5}}
%\end{figure}

\subsection{Region 6 with $C\cdot G$} \label{sec:region6}

\noindent Once $A = (b/a)^{1/2}$, the $\overline{T_1}$ point has reached $w = \infty$ and for larger values of $A$, the turning point begins to move from right-to-left along the downstream solid boundary. 

There is one issue which needs to be addressed: in Figure \ref{fig:gcwaves_region456}(c), the \Mycirct{B} $>$ \Mycirct{G} line intersects the \Mycirct{G} $>$ \Mycirct{C} line at a \emph{Stokes Crossing Point} or SCP. If we analytically continue in a path that encircles the SCP, then we find that in order to avoid an inconsistency, the base solution must switch-off a capillary wave somewhere along the path; in other words, there must be a \Mycirct{B} $>$ \Mycirct{C} line, which goes through the SCP. However, neither the corner, nor the stagnation point, produces such a Stokes line.

This issue is addressed in Appendix \ref{sec:crossing}, where we discuss three key ideas: (i) there is a second singularity [marked by the black circle in Figure \ref{fig:gcwaves_region456}(c)]. This singularity does not appear in any of the early terms \eqref{eq:gcwave_theta0}--\eqref{eq:gcwave_q1}, but represents a singularity in the late terms \emph{of} the late terms. It lies on the adjacent Riemann sheet from the originally defined corner point, $\zeta = a$, and can be detected by setting $\chi = 0$, where from \eqref{eq:gcwave_chiprime}
\[
 \chi = -i \int_\Gamma \left[ \frac{q_0^2 -
\sqrt{\Delta}}{2\tau q_0} \right] \ d\varphi.
\] 

\noindent The contour $\Gamma$ begins at $\zeta = a$, crosses the branch cut from the turning point, $\underline{T_1}$, and continues along the secondary Riemann sheet of $\smash{\sqrt{\Delta}}$. 

In the appendix, we also establish that (ii) the secondary singularity produces a \Mycirct{B} $>$ \Mycirct{C} Stokes line that crosses the SCP [shown dashed in Figure \ref{fig:gcwaves_region456}(c)]. Finally, (iii) the portion of the \Mycirct{B} $>$ \Mycirct{C} line that joins the SCP to the singularity must be inactive; this switching-off of a Stokes line is due to the higher-order Stokes Phenomenon. These more complicated issues of secondary singularities, Stokes crossing points, and the higher-order Stokes Phenomenon, have been studied by others in a variety of different situations [see for example \cite{berk_1982, howls_2004, daalhuis_2004, chapman_1999b, chapman_2005, chapman_2007}].

The solution is then
\begin{equation*}
\left(-\mathcal{C} + \mathcal{G}_{\underline{T_1}}\right) e^{X_\text{cap}}
\Swtch{Stag.}{B}{C} 
\mathcal{G}_{\underline{T_1}} e^{X_\text{cap}}
\Swtch{Sec. Sing.}{B}{C}
0 
\Swtch{Corn.}{B}{G}
\mathcal{G} e^{X_\text{grav}},
\end{equation*}

\noindent where the values of $\mathcal{C}$ and $\mathcal{G}$ are given by \eqref{eq:gcwave_region1_AC}. The pre-factor $\mathcal{G}_{\underline{T_1}}$ reminds us that the capillary wave is switched-on by taking the $\mathcal{G}$ gravity wave and crossing the Stokes line from the the turning point; its value can be computed from \eqref{eq:gcwave_turningexp}. These doubly exponentially small capillary waves are switched off downstream upon crossing the Stokes line from the secondary singularity.

Decaying waves on either side of the step are exactly what we expect from the linear analysis; however, we see that the result is not quite the same as in the linearised theory due to the presence of the doubly-small capillary waves upstream. The free surface is shown in Figure \ref{fig:gcwaves_region456}(c). However, we note that, as for the case of pure gravity waves (see \citealt{chapman_2006}), we would expect a distinguished limit in the small Froude, small Bond, and small step height limits. Thus relating the results here with the work in Part 1 is not likely to be a trivial problem. 

\section{Discussion}

\noindent Now having reached the end of our work on the gravity-capillary problem, we have one final query: \emph{do these new waves truly exist?}

Our theoretical predictions form the first attempt of exploring the
previously unknown region of low-Froude and low-Bond numbers. Our results posit the existence of six different families of waves for flow over a step (\ding{182} to \ding{187} in Figure \ref{fig:gcwave_betatau}), and reveals that the usual dispersion curve separating linear solutions \ding{182} and \ding{187} widens if we include the nonlinear nature of the geometry. While this structure is only valid at low Froude and low Bond numbers, it also sheds new light on the general complexity of gravity-capillary problem, which has been freely acknowledged in the past.

Many open questions remain, some of which may ultimately hold the key to detecting these new waves (or discounting their existence). Throughout this work, we have provided a clear analysis of the \emph{local} issues of the gravity-capillary problem. For example, we have shown how the emergence of a Stokes line depends on the local angle of the obstruction, or how the switching-on of waves near turning points can be predicted via an Airy-like transition. Although we have not shown the tedious calculations, the crucial pre-factor, $\Lambda$, is another local issue, and can be derived by matching inner and outer solutions near the singularities (see \citealt{trinh_thesis} for more details). But what about the \emph{global} properties of the Stokes lines? 

For example, can a Stokes line emerge from a singularity along a secondary Riemann sheet, only to later return to the primary sheet and intersect the free surface? Our analysis in Appendix \ref{sec:crossing} attempted to answer this question by drawing a comparison to a similar problem containing multiple singularities and crossing Stokes lines; we then used these results to propose the likely free-surface configuration for the complicated situation of Region 6 of Figure \ref{fig:gcwaves_region456}(c). However, a more rigorous treatment of these issues remains an open problem. 

Equally fascinating is the myriad of configurations that now seems possible for free-surface gravity-capillary flows over different geometries. A classic question is to inquire whether there are special geometries for which leading-order wave cancellation occurs and where, for example, the waves produced by one singularity cancels the waves produced by another. Although these issues regarding variable geometries have been studied in the context of pure gravity or capillary waves (\citealt{chapman_2002, trinh_1hull, lustri_2012, trinh_nhull}), a catalogue of the wave configurations for different geometries in the combined gravity-capillary case is a subject of future work. 

Finally, the most significant question is why have these waves not been previously detected? From the perspective of numerical simulations, one plausible difficulty concerns the so-called \emph{radiation problem}: it is unclear how the upstream radiation condition can be accurately implemented in a numerical scheme. This difficulty is highlighted in, for example, the works of \cite{forbes_1983}, \cite{scullen_thesis}, and \cite{grandison_2006}. For gravity-capillary flow past a large obstruction, the resultant solutions are influenced by the chosen treatment of the far field, and it is possible that inherent errors in current numerical methods make it difficult to differentiate between the structure of our new gravity-capillary waves and the typical linearised solutions. Do these waves truly exist? Our theoretical results suggest that they do; the search for such configurations, either in nature or in the digital world, forms the basis of ongoing investigation.

% The most important issue for future work, then, is the search for such waves, in nature or in the digital world.

% Our theoretical results suggest that they do, and we hope the community will join us in our continued search for their presence. 

\appendix

\section{On the crossing of Stokes lines} \label{sec:crossing}

\noindent We address the issue of the crossing of Stokes lines in the gravity-capillary problem by drawing an analogy to the same phenomenon which occurs in a modified Airy equation:
\[
 \epsilon^2 y'' = xy + \frac{e^{x/\epsilon}}{x-a}.
\]

\noindent This equation was chosen because it not only possesses an (Airy) turning point at $x = 0$, but also an additional singularity at $x = a$. Writing $y = ve^{x/\epsilon}$ gives
\begin{equation} \label{eq:modairy_veq}
 \epsilon^2 v'' + 2\epsilon v' + (1-x) v = \frac{1}{x - a},
\end{equation}

\noindent where we also require $v \to 0$ as $|x| \to \infty$. We can now perform the standard asymptotic analysis of \eqref{eq:modairy_veq}, expanding $v = \sum \epsilon^n v_n$, and giving the first two orders as
\begin{alignat}{3}
v_0 &= \frac{1}{(1-x)(x-a)} \label{eq:modairy_v0} \\ 
v_1 &= \frac{-2}{(a-x)^3(x-a)} + \frac{2}{(1-x)^2(x-a)^2}, \label{eq:modairy_v1}
\intertext{and in general, for $n \geq 2$,}
v_n &= - \frac{1}{1-x} \biggl[ v_{n-2}'' + 2v_{n-1}' \biggr]. \label{eq:modairy_vn}
\end{alignat}

\noindent As $n \to \infty$, we expect the singularities at $x = 0$ and $x = a$ to produce factorial over power divergence of the late terms. Substituting the ansatz
\[
 v_n \sim \frac{V(x) \Gamma(n+\gamma)}{[\chi(x)]^{n+\gamma}},
\]

\noindent into \eqref{eq:modairy_vn} gives $(\chi')^2 - 2\chi' + (1-x) = 0$, or solving yields
\begin{equation} \label{eq:modairy_chigen}
 \chi = \int^x 1 \pm s^{1/2} \ ds. 
\end{equation}

\noindent There are clearly singularities at $x = 1$ and $x = a$, and we shall focus on the exponentials generated by the latter. At a Stokes line emerging from the singularity at $x = a$, the base series can switch-on exponentials of the form, $e^{-\chi/\epsilon}$ with $\chi$ being one of
\begin{align}
  \chi_1 &= x + \frac{2}{3}x^{3/2} - a - \frac{2}{3} a^{3/2} \label{eq:modairy_chi1} \\
  \chi_2 &= x - \frac{2}{3}x^{3/2} - a + \frac{2}{3} a^{3/2}. \label{eq:modairy_chi2}
\end{align}

\noindent If we use \Mycirct{B} to denote the base series, and \Mycirct{1} and \Mycirct{2} for the two exponentials, then the Stokes Phenomenon describes the process in which \Mycirct{B} $>$ \Mycirct{1} or \Mycirct{B} $>$ \Mycirct{2}. However, there is also a turning point at $x = 0$, for which the associated Stokes lines can cause interactions between the two exponentials. If \Mycirct{1} $>$ \Mycirct{2}, then this produces an exponential with 
\begin{equation}
  \chi_3 = x - \frac{2}{3} x^{3/2} - a - \frac{2}{3}a^{3/2} \label{eq:modairy_chi3},
\end{equation}

\noindent since the Airy transition simply requires switching the branch of $x^{3/2}$. Similarly, if \Mycirct{2} $>$ \Mycirct{1}, this produces an exponential with
\begin{equation}
  \chi_4 = x + \frac{2}{3} x^{3/2} - a + \frac{2}{3}a^{3/2}. \label{eq:modairy_chi4}
\end{equation}

\begin{figure} \centering
\includegraphics[width=0.49\textwidth]{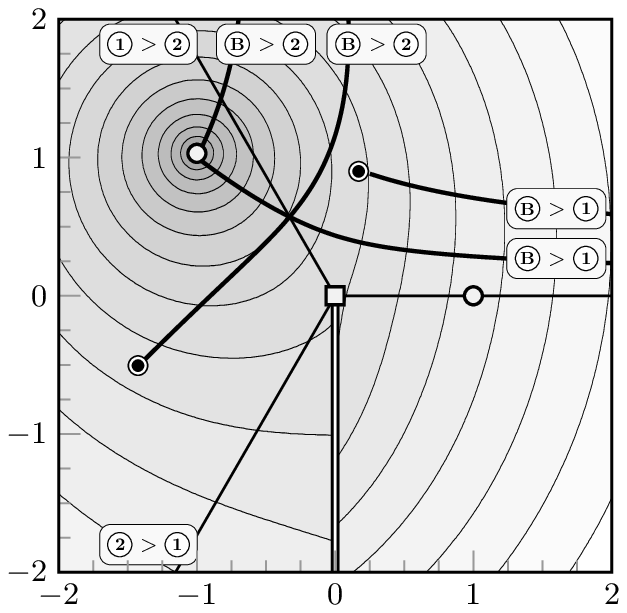}
\includegraphics[width=0.49\textwidth]{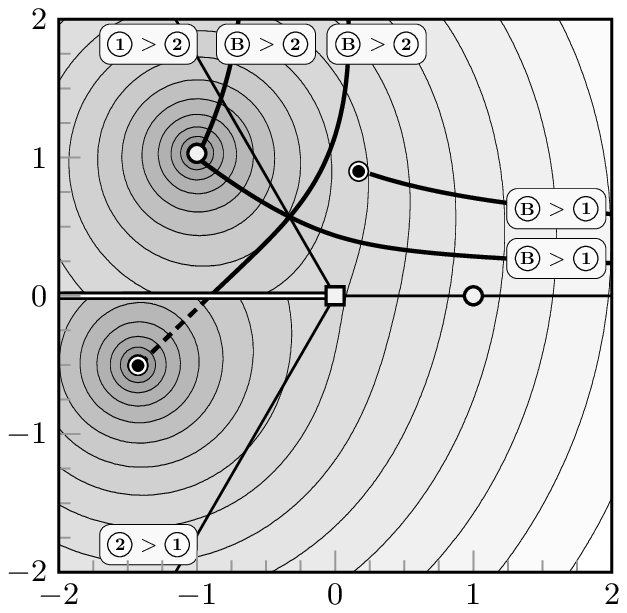}
  \caption{Stokes lines in the complex $x$-plane, shown with two different choices of branch cuts (left/right). White circles denote the singularities at $x = a$ and $x= 1$, black circles are the secondary singularities, and a square denotes the turning point. The Stokes lines from $x = 1$ are not drawn. Branch cuts are shown as a striped line, and the dashed lines are used to indicate points on a secondary Riemann sheet. Background contours are for $|\chi_3|$ (left) or $|\chi_1|$ (right), with darker regions indicating smaller values. \label{fig:modairy}} 
\end{figure}

\noindent The Stokes lines for the case $a = -1 + i$ are shown in the left frame of Figure \ref{fig:modairy}. Notice that the \Mycirct{B} $>$ \Mycirct{1} Stokes line intersects the \Mycirct{1} $>$ \Mycirct{2} line at a Stokes Crossing Point (SCP). If we analytically continue in a circle around the SCP, we see that a third Stokes line, with \Mycirct{B} $>$ \Mycirct{2}, is needed in order to avoid an inconsistency. From \eqref{eq:modairy_chi3}, this seems to require a singularity at the point $x = x_*$, where $\chi_3 = 0$; for $a = -1 + i$, the singularity lies at $x_* \approx -1.423 - 0.509i$. There is a similar singularity at $x \approx 0.249 + 0.883i$, where $\chi_4 = 0$, which we associate with a \Mycirct{B} $>$ \Mycirct{1} line. The curiosity, however, is that these singularities do not appear anywhere in the base series, \Mycirct{B}, and so they do not seem to be associated with any eventual divergence.

To explore this in more detail, we define the Fourier transform as 
\[
 \widehat{v}(k) = \frac{1}{\sqrt{2\pi}} \int_{-\infty}^\infty v(x) e^{ikx} \ dx,
\]

\noindent and take the transform of \eqref{eq:modairy_veq}, giving 
\[
 i\epsilon^2 k^2 \widehat{v} - 2\epsilon k \widehat{v} - i\widehat{v} + \frac{d\widehat{v}}{dk} = \sqrt{2\pi} e^{iak} H(k),
\]

\noindent where $H(k)$ is the Heaviside function, and we have assumed that $\Im(a) > 0$. Solving this equation yields
\[
\widehat{v} = \sqrt{2\pi} \exp \left[ \frac{i\epsilon^2 k^3}{3} + \epsilon k^2 + ik\right] 
\int_0^k \exp \left[ \frac{i\epsilon^2 u^3}{3} - \epsilon u^2 - iu + iau \right] du,
\]

\noindent or once inverted,
\[
 v =  \int_{-\infty}^\infty \exp \left[ \frac{i\epsilon^2 k^3}{3} + \epsilon k^2 + ik - ikx \right] 
\int_0^k \exp \left[ \frac{i\epsilon^2 u^3}{3} - \epsilon u^2 - iu + iau \right] du \, dk.
\]

\noindent We rescale $k = s/\epsilon$ and $i = v/\epsilon$, giving
\begin{equation} \label{eq:airy_vint}
 v = \int_{-\infty}^\infty \exp \left[ - \frac{\phi(s; x)}{\epsilon} \right] 
\int_0^s \exp \left[ \frac{\phi(v; a)}{\epsilon} \right] dv \, ds,
\end{equation}

\noindent where we have defined
\[
 \phi(v; a) = \frac{iv^3}{3} - v^2 - iv + aiv.
\]

\noindent In the limit $\epsilon \to 0$, the dominant contributions of the innermost integral in \eqref{eq:airy_vint} come from the end points at $v = 0$ and $v = s$, or at the two saddle points at $v = -i \pm i\sqrt{a}$. Thus \eqref{eq:airy_vint} can be approximated as a sum of integrals of the form
\begin{equation} \label{eq:modairy_intA}
 \int_{-\infty}^\infty A(s) e^{\psi(s)/\epsilon} \ ds,
\end{equation}

\noindent where $\psi$ is one of 
\begin{numcases}{\psi(s) = }
is(a -x) & \label{eq:modairy_psi1} \\ 
-\frac{is^3}{3} + s^2 + is - isx & \label{eq:modairy_psi2} \\ 
-\frac{is^3}{3} + s^2 + is - isx - \frac{1}{3} + a \pm \frac{2a^{3/2}}{3} & \label{eq:modairy_psi3}
\end{numcases}

\noindent Approximating the integral \eqref{eq:modairy_intA} using \eqref{eq:modairy_psi1} recovers the base series \eqref{eq:modairy_v0}--\eqref{eq:modairy_v1}. The second and third expressions for $\psi$ have saddle points at $s = -i \pm i \sqrt{x}$. Using the second expression \eqref{eq:modairy_psi2}, we see that the integral \eqref{eq:modairy_intA} produces contributions $e^{-\chi/\epsilon}$ with 
\[
 \chi = -\frac{1}{3}  + x \pm \frac{2}{3} x^{3/2}. 
\]

\noindent One of these comes from the singularity at $x = 1$; the other comes from the same singularity, and then going around the turning point. 

Finally, approximating the integral at the saddle points using the third expression for $\psi$ in \eqref{eq:modairy_psi3}, gives four possible expressions for $\chi$. Two of these are directly from the singularity at $x = a$, and return $\chi_1$ and $\chi_2$ from \eqref{eq:modairy_chi1}--\eqref{eq:modairy_chi2}. The other two, returning $\chi_3$ and $\chi_4$ in \eqref{eq:modairy_chi3}--\eqref{eq:modairy_chi4}, are generated by the same singularity, but involve an integration contour which goes around the turning point. This verifies that the missing Stokes line should be there, though it does not shed much light on the mysterious singularity, $x = x_*$. 

Let us go back to the expansion. We can write the $n^\text{th}$ term, $v_n$, from \eqref{eq:modairy_vn} as
\begin{equation} \label{eq:modairy_vnseries}
 v_n = \sum_{k=1}^{n+1} \sum_{m=0}^{2n} \frac{a_{k,m}}{(x-a)^k (1-x)^m},
\end{equation}

\noindent with $a_{k,m}$ given by
\begin{multline*}
 a_{k,m} = 2(k-1)a_{k-1,m-1} - 2(m-2)a_{k,m-2} - (k-2) a_{k-2,m-1} - (k-2)^2 a_{k-2,m-1} \\ 
 + 2(k-1)(m-2)a_{k-1,m-2} - (m-3)a_{k,m-3} - (m-3)^2 a_{k,m-3}.
\end{multline*}

\noindent We believe that if this recurrence relation is solved, and the method of steepest descents is used to approximate the terms \eqref{eq:modairy_vnseries} as $n \to \infty$, then the mysterious point, $x = x_*$, would appear as a singularity in the late terms \emph{of} the late terms, $v_n$. These `secondary' singularities and SCPs have been encountered in \cite{chapman_2005} and \cite{howls_2004}.

Before moving to the gravity-capillary problem, we clarify a labelling issue: the prescription of $x_*$ by $\chi_3 = 0$ in \eqref{eq:modairy_chi3} is correct only if the \Mycirct{B} $>$ \Mycirct{2}\, Stokes line can be continued from the SCP to the secondary singularity, $x_*$, along the same branch of $x^{3/2}$ (as it is drawn in the left frame of Figure \ref{fig:modairy}). However, a different choice of the branch cut (as in the right frame) shows that $\chi_1 = 0$ is the correct equation. To correct this ambiguity, we use an alternative notation. Let $x_{(k)}$ correspond to a point on the $k^\text{th}$ Riemann sheet (associated with the two branches of $x^{3/2}$); that is, $x_{(k)}$ is mapped to $e^{\pi i k}x^{3/2}$, with $k = 0, 1$. From \eqref{eq:modairy_chigen}, we then write $\chi_1$ as 
\begin{equation} \label{eq:modairy_chi1alt}
 \chi_1 = \int_{a_{(0)}}^x \Bigl[1 + e^{\pi i k(s)}s^{3/2} \Bigr] \ ds, 
\end{equation}

\noindent where $k(s) = 0$ or $1$, depending on which sheet the integrand is currently on. This makes it very clear that the secondary singularity, $x_*$, is still given by $\chi_1 = 0$, but $x_*$ may either lie on the sheet with $a_{(0)}$, or the sheet with $a_{(1)}$. Thus, $\chi_3$ in \eqref{eq:modairy_chi3} is simply 
\[
 \chi_3 = \left(\int_{a_{(0)}}^{a_{(1)}} + \int_{a_{(1)}}^{x}\right) \Bigl[1 + e^{\pi i k(s)}s^{3/2} \Bigr] \ ds, 
\]

\noindent except that $x$ is restricted to the same sheet as $a_{(1)}$, since \eqref{eq:modairy_chi3} requires principal branches taken throughout. Writing $\chi_1$ in the form \eqref{eq:modairy_chi1alt} serves to emphasize the fact that the contribution $\chi_3$ is found by beginning at $a_{(0)}$ and then crossing onto the secondary sheet. 

\subsection{Crossing Stokes lines in the gravity-capillary problem}

We now address the issue of the crossing Stokes lines in Region 6 of \S\ref{sec:region6}. For $\chi$ in \eqref{eq:gcwave_chi}, there are three types of branch points, but only the ones from the turning points are relevant. Remembering that the corner of the step is at $\zeta = a$, and changing to integration in $\zeta$ using \eqref{wzeta}, we denote, for the gravity wave,
\begin{equation} \label{eq:gcwave_chig1}
\chi_g = i \int_{a_{(1)}}^\zeta \left[ \frac{q_0^2 + e^{\pi i k(s) }\sqrt{\Delta}}{2\tau q_0} \right] \frac{1}{s} \ ds,
\end{equation}

\noindent where $k(s) = 1$ on the branch of $a_{(1)}$ and $k(s) = 0$ on the branch of $a_{(0)}$ (associated with the capillary wave). In addition to the two turning points, there is a logarithmic branch point from the factor of $s^{-1}$, and two square-root branch points from the two singularities of $q_0$. The former type adds a constant to $\chi$ upon crossing the cut, while latter type switches $q_0$ to $-q_0$ upon crossing the cut. If we wanted to fully explore the different sheets, it would be better to write $\zeta = \zeta_{(k_1, k_2, k_3, k_4, k_5)}$, to keep track of how many times we have gone around each of the five branch points.  However for our purpose, only the single turning point which causes the intersection is important.

\begin{figure} 
\includegraphics[width=1.0\textwidth]{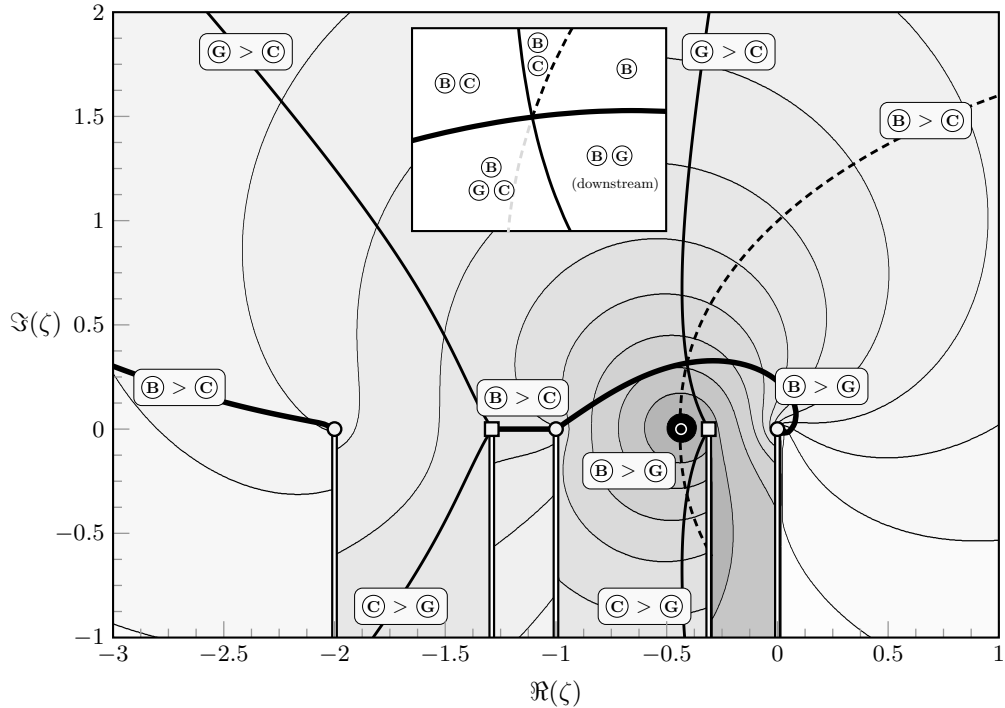}
  \caption{%
  The underlying contours and shading represent $|\chi_g|$ from \eqref{eq:gcwave_chig2}, with dark regions indicating small values. Values of $\beta = 1$ and $\tau = 1.5$ were used, and $a = 1$, $b = 2$ for the corner and stagnation points (circles). Turning points are $\zeta_1 \approx -0.310$ and $\zeta_2 \approx -1.290$ (squares), and the Stokes lines were computed using \eqref{eq:gcwave_chistokes} and \eqref{eq:gcwave_tpstokes}. Note that with the exception of the Stokes line from $\zeta_2$, all the Stokes lines that begin in the upper-half plane will eventually intersect $\zeta\in\mathbb{R}^+$. Branch cuts are shown striped. The secondary singularity, located at $\zeta^\bullet \approx -0.435$, is shown as a black circle, and its Stokes line shown dashed (note that this Stokes line, as well as the singularity, lies on a different Riemann sheet than the other components of the figure). The inset confirms that if we analytically continue around the SCP, beginning with the base solution upstream, then a portion of the new Stokes line must switch off across the SCP (gray, dashed).   \label{fig:gcwave_realcont}} 
\end{figure}

Remember: the issue is that the crossing of the \Mycirct{B} $>$ \Mycirct{G} and \Mycirct{G} $>$ \Mycirct{C} lines in Figure \ref{fig:gcwaves_region456}(c) also requires a \Mycirct{B} $>$ \Mycirct{C} line in order to avoid an inconsistency. Like the example of the preceding section, this missing Stokes line comes from a secondary singularity, $\zeta = \zeta^\bullet$, not present in any of the early orders. To see this, we write \eqref{eq:gcwave_chig1} in the form 
\begin{equation}\label{eq:gcwave_chig2}
\chi_g = i\left( \int_{a_{(1)}}^{a_{(0)}} +  \int_{a_{(0)}}^\zeta \right) \left[ \frac{q_0^2 + e^{\pi i k(s) }\sqrt{\Delta}}{2\tau q_0} \right] \frac{1}{s} \ ds,
\end{equation}

\noindent and we further restrict $\zeta$ to the $k = 0^\text{th}$ Riemann sheet. Thus, \eqref{eq:gcwave_chig2} is an integral from $a_{(1)}$ to $a_{(0)}$ via the branch cut from the turning point, and then proceeds from $a_{(0)}$ to $\zeta$. The second integral is the usual capillary integral (hence the capillary wave). This new representation is used to make Figure \ref{fig:gcwave_realcont}; the most important feature of the figure is the missing singularity and its missing Stokes line, which indeed intersects the SCP. 

In addition to the appearance of the secondary singularity, intersecting Stokes lines are typically accompanied by yet another subtlety: that of the Higher-Order Stokes Phenomenon [see again \cite{howls_2004} and \cite{chapman_2005}]. This phenomenon specifies that at a SCP, Stokes lines may themselves switch-off. Indeed, this must be the case because if we analytically continue around the SCP in the inset of Figure \ref{fig:gcwaves_region456}(c), we see that both portions of the \Mycirct{B} $>$ \Mycirct{C} line (on either side of the SCP) can not be active. It remains to determine whether the portion of the Stokes line connecting the singularity to the SCP is active, or whether it is rather the portion from the SCP to the free surface. 

To address this issue, we note in Figure \ref{fig:gcwave_realcont} that if the dashed Stokes line is followed from the SCP to the secondary singularity into the lower half-plane, then it changes from a \Mycirct{B} $>$ \Mycirct{C} Stokes line to a \Mycirct{B} $>$ \Mycirct{G} Stokes line across the secondary singularity ($\zeta = \zeta^\bullet$). In fact, the portion of the axis where $-b \leq \zeta \leq \zeta_1$, corresponds to the Anti-Stokes lines (where \Mycirct{B} $=$ \Mycirct{C} $=$ \Mycirct{G}) from the corner ($\zeta = -b$) and turning point ($\zeta = \zeta_1$). Across the secondary singularity, then, the Stokes line has changed character abruptly, and we argue by analogy to studies of the Higher-Order Stokes Phenomenon that this can not occur unless $\zeta^\bullet$ is a turning point (which it is not). Thus, the portion of the Stokes lines connected to the secondary singularity is inactive. Finally, we add that in Figure \ref{fig:gcwave_realcont}, the dashed Stokes line is the only line associated with secondary Riemann sheets that we have chosen to show. There are other Stokes lines we have not shown (and in fact, other secondary singularities), but these are not relevant to the free surface waves.

%\bibliographystyle{/u/ptrinh/work/documents/bib/styles/jfm}
%\bibliography{/u/ptrinh/work/documents/bib/philmaster}
%\bibliographystyle{/Volumes/InternalHD/Users/trinh/work/documents/bib/styles/jfm}
%\bibliography{/Volumes/InternalHD/Users/trinh/work/documents/bib/philmaster}
\bibliographystyle{jfm}
\providecommand{\noopsort}[1]{}

\end{document}